\documentclass[12pt,preprint]{aastex}
\shorttitle{Associated Mg II absorption systems}
\shortauthors{Vanden Berk et al.}

\def \nh{\ensuremath{{\rm N}_{\rm H\;I}}}
\def \wmg{\ensuremath{{\rm W}_{\rm Mg\;II}}}

\def \wc4{\ensuremath{{\rm W}_{\rm C\;IV}}}
\def \kms{\ensuremath{{\rm km\;s}^{-1}}}
\def \zem{\ensuremath{z_{em}}}
\def \zab{\ensuremath{z_{abs}}}
\def \ebv{\ensuremath{E(B-V)}}
\def \bet{\ensuremath{\beta}}
\def \imag{\ensuremath{i\;{\rm magnitude}}}
\def \dgi{\ensuremath{\Delta(g-i)}}
\def \wmgr{\ensuremath{{\rm W}_{\rm Mg\;I}/{\rm W}_{\rm Mg\;II}}}
\def \walr{\ensuremath{{\rm W}_{\rm Al\;II}/{\rm W}_{\rm Al\;III}}}
\shorttitle{Associated {Mg~II} II absorbers}
\shortauthors{Vanden Berk et al.}
\begin{document}
\title{Average Properties of a Large Sample of $z_{abs}\sim z_{em}$ associated
{Mg~II} Absorption Line Systems}
\date{}
\author{Daniel Vanden Berk}\affil{Department of Astronomy and Astrophysics, The
Pennsylvania state University, 525 Davey Laboratory, University Park, PA 16802,
USA} \author{Pushpa Khare}\affil{Department of Physics, Utkal University,
Bhubaneswar, 751004, India} \author{Donald G. York\altaffilmark{1}}\affil{Department of
Astronomy and Astrophysics, University of Chicago, Chicago, IL 60637,
USA} \author{Gordon T. Richards\altaffilmark{2}}\affil{Department of Physics, Drexel
University, 3141 Chestnut Street, Philadelphia, PA 19104,
USA}\author{Britt
Lundgren}\affil{Department of Astronomy, University of Illinois at
Urbana-Champaign, MC-221 1002 W. Green Street, Urbana, IL 61801, USA} \author{Yusra
Alsayyad}\affil{Department of
Astronomy and Astrophysics, University of Chicago, Chicago, IL 60637,
USA} \author{Varsha P. Kulkarni}\affil{Department of Physics and
Astronomy, University of South Carolina, Columbia, SC 29207, USA} \author{Mark
SubbaRao\altaffilmark{3}}\affil{Department of
Astronomy and Astrophysics, University of Chicago, Chicago, IL 60637,
USA} \author{Donald  P. Schneider}\affil{Department of Astronomy and Astrophysics, The
Pennsylvania state University, 525 Davey Laboratory, University Park, PA 16802,
USA}  \author{Tim
Heckman}\affil{Department of Physics and Astronomy, Homewood Campus,
Baltimore, MD 21218, USA} \author{Scott Anderson}\affil{Department of Astronomy,
University of Washington, Box 351580, Seattle, WA 98195, USA} \author{Arlin P. S.
Crotts}\affil{Department of Astronomy, Columbia University, New York, NY
10027, USA}  \author{Josh Frieman and C. Stoughton}\affil{Fermilab National Accelerator
Laboratory, P.O. Box, 500, Batavia, IL, 60510, USA} \author{James
T.  Lauroesch\altaffilmark{4}}\affil{Department of Physics and Astronomy, University of
Louisville, Louisville, KY 40292} \author{Patrick B.
Hall}\affil{Department of Physics and Astronomy, York University, 4700
Keele Street, Toronto, ON M3J 1P3, Canada} \author{Avery
Meiksin}\affil{Institute of Astronomy, Royal Observatory, University of
Edinburgh, Blackford hill, Edinburg, EH9 3HJ, U.K.} \author{Michael Steffing}\affil{Department of
Astronomy and Astrophysics, University of Chicago, Chicago, IL 60637,
USA} 
\author{Johnny Vanlandingham}\affil{Department of
Astronomy and Astrophysics, University of Chicago, Chicago, IL 60637,
USA} 
\altaffiltext{1}{Also, Enrico Fermi Institute, University of Chicago, Chicago,
IL 60637, USA}
\altaffiltext{2}{Also, Department of Physics and Astronomy, The Johns Hopkins
University, 3400 North Charles Street, Baltimore, MD 21218, USA}
\altaffiltext{3}{Also, Adler Planetarium and Astronomy Museum, 1300 S.  Lake
Shore Drive, Chicago, IL 60605, USA}
\altaffiltext{4}{Also, Department of Physics and Astronomy, Northwestern
University, Evanston, IL 60208, USA}

\begin{abstract} We have studied a sample of 415 associated ($z_{abs}\sim
z_{em}$; relative velocity with respect to QSO $<$ 3000 \kms) {Mg~II}
absorption systems with 1.0$\le z_{abs}\le$1.86, in the spectra of Sloan
Digital Sky Survey, Data Release 3 QSOs, to determine the dust content,
ionization state, and relative abundances in the absorbers.  We studied the
dependence of these properties on the properties of the backlighting QSOs and
also, compared the properties with those of a similarly selected sample of 809
intervening systems (apparent relative velocity with respect to the QSO of $>$
3000 \kms), so as to understand their origin.  Normalized, composite spectra
were derived, for absorption line measurements, for the full sample and for
several sub-samples, chosen on the basis of the line strengths and other
absorber and QSO properties.  Composite absorption lines differ in small but
measurable ways from those in the composite spectra of intervening absorption
line systems, especially in the relative strengths of {Si~IV}, {C~IV} and
{Mg~II}. From the analysis of the composite spectra, as well as from the
comparison of measured equivalent widths in individual spectra, we conclude
that the associated {Mg~II} absorbers have higher apparent ionization, measured
by the strength of the {C~IV} absorption lines compared to the {Mg~II}
absorption lines, than the intervening absorbers.  The ionization so measured
appears to be related to apparent ejection velocity, being lower as the
apparent ejection velocity is more and more positive. Average extinction curves
were obtained for the sub-samples by comparing their geometric mean QSO spectra
with those of matching (in \zem $~$and \imag) samples of QSOs without
absorption lines in their spectra.  There is clear evidence for dust-like
attenuation in these systems, though the 2175 {\AA} absorption feature is not
present: the extinction is similar to that found in the Small Magellanic Cloud.
The extinction is almost twice that observed in the similarly selected sample
of intervening systems. We reconfirm with our technique that QSOs with non-zero
FIRST (Faint Images of the Radio Sky at Twenty-cm) radio flux are intrinsically
redder than the QSOs with no detection in the FIRST survey. The incidence of
associated {Mg~II} systems in QSOs with non-zero FIRST radio flux is 1.7 times
that in the QSOs with no detection in the FIRST survey. The associated
absorbers in radio-detected QSOs which comprise about 12\% of our sample, cause
three times more reddening than the associated absorbers in radio-undetected
QSOs. The origin of this excess reddening in the absorbers is indicated by the
correlation of the reddening with the strength of Mg II absorption. This excess
reddening possibly suggests an intrinsic nature for the associated absorbers in
radio-detected QSOs. 
\end{abstract}
\section{Introduction} Many of the first detected, narrow-line QSO absorption
systems had redshifts close to those of their QSOs (e.g. Stockton \& Lynds
1966; Burbidge, Lynds \& Burbidge 1966). Such systems, having a relative
velocity with respect to the QSO, in units of the speed of
light\footnote{$\bet={{(1+\zem)^2-(1+\zab)^2}\over{(1+\zem)^2+(1+\zab)^2}}$},
$\beta$, smaller than 0.02, are now termed associated systems. Study of the
associated systems is important from the point of view of understanding the
energetics and kinematics near the central black hole and also for
understanding the ionization structure, dust content and abundances in material
directly exposed to the radiation from the QSOs, and in some cases, possibly
ejected from, the QSOs or the accretion disks. Associated systems have been
suggested to arise (a) in the outer parts of QSO host galaxies (e.g. Heckman
et al. 1991; Chelouche et al.  2007) which may have gas properties similar to
those in the outer parts of inactive galaxies (Steidel et al. 1997, Fukugita
\& Peebles 2006); (b) by material within 30 kpc of the AGN, accelerated by
starburst shocks from the inner galaxy (e.g. Heckman et al. 1990, 1996;
D'Odorico et al.  2004; Fu and Stockton 2007a); (c) in the core of the AGN,
within 10 pc of the black hole (Hamann et al. 1997a; Hamann et al. 1997b;
Barlow \& Sargent 1997). 

In case (a), these are possibly ``halo" clouds as in normal galaxies, possibly
lit up by the QSO to produce extended regions of Lyman alpha emission, but
not thought to be moving at the high velocities necessary to explain the
dispersion of associated system velocities with respect to the QSO itself.
However, using spectra of a set of angularly close QSO pairs, Bowen et al.
(2006) confirmed that gas in the outer parts of some QSO host galaxies is
detectable as absorption in the spectra of background QSOs, but the spectra of
the foreground QSOs did not reveal associated absorption systems: evidently,
the appearance of the absorption is dependent on the angle of the line of sight
to the spin axis of the accretion disk. The low velocities expected from such
gas led to the postulate that clusters of galaxies near the QSO showing
associated absorption were responsible for the absorption and the dispersion of
cloud velocities.  However, QSOs are not only associated with clusters of
galaxies: they appear in a wide range of galaxy types and masses (Jahnke et
al. 2004) and, while found in slightly higher density environments (Serber et
al. 2006), do not require cluster type densities on large scales (Wake et al.
2004).
 
In case (b), the gas is typically found to have densities of a few hundred
particles cm$^{-3}$, high enough to produce excited fine structure lines of
{Si~II} and {C~II}.  There are suggestions in the above references that this is
material ejected by a galactic wind,  possibly a superwind from a starburst
(Heckman et al. 1990), then lit up by the QSO. (These are often called extended
emission line regions, EELR). Cooling flows (Crawford \& Fabian 1989) no longer
seem to be considered in most cases (Fu \& Stockton 2007b).
 
In case (c),  absorption is inferred to be very close to the black hole because
of variability in the absorption lines, and/or because of the presence of
clouds that do not cover the source. Theoretical investigations of case (c)
(Arav et al. 1994; Konigl \& Kartje 1994; Murray et al. 1995; Krolik \& Kriss
2001; Proga et al. 2000; Everett 2005; Chelouche \& Netzger 2005 (for a Seyfert
galaxy with a much lower radiation field)),  have reinforced the plausibility
of thermal or hydromagnetic, radiation assisted flows, both parallel to the AGN
jet axis, and perpendicular to it, along the accretion disk. See, for example,
Figure 13 of Konigl \& Kartje (1994), Figure 1 of Richards et al.  (1999),
Figure 1 of Murray et al.  (1995) and Figure 1 of Everett (2005).

Early studies concentrated on systems with C IV doublets. Weymann et al. (1979)
and Foltz et al. (1986) found a statistical excess of such systems as compared
to what was expected if these were randomly distributed in space. Other studies
(Young et al. 1982; Sargent et al. 1988) could not confirm these observations.
Later it was shown that {Mg~II} systems (Aldcroft et al. 1994) and C IV systems
(Anderson et al. 1987; Foltz et al. 1988) with \bet $<$ 0.0167 are
preferentially found in steep-spectrum radio sources, often thought of as
sources dominated by emission from lobes rather than the core of the source.
Ganguly et al. (2001) showed that high ionization systems (having lines of C
IV, N V and O VI) with \bet $<$ 0.0167 with $z_{abs}<1.0$ are not present in
radio-loud QSOs that have compact radio morphologies, flat radio spectra (core
dominated sources) and  C IV lines with mediocre FWHM ($\le 6000$ \kms). Baker
et al.  (2002) from a study of a near-complete sample of low-frequency
selected, radio-loud QSOs corroborated trends for C IV associated absorption to
be found preferentially in steep-spectrum and lobe-dominated QSOs, suggesting
that the absorption is the  result of post-starburst activity and that the C IV
lines weaken as  the radio source grows, clearing out the gas and dust.
Vestergaard (2003, hereafter V03) studied a sample of high ionization
associated systems with $1.5<z_{abs}<3.5$ and found that the occurrence, or
not, of such systems is independent of radio properties of the QSOs. V03 did
not find the relationship between radio source  size and C IV line strength of
Baker et al.  (2002), nor the absence of absorbers in low line width systems of
Ganguly et al (2001). V03 did identify weak correlations among QSO properties
and the line strengths of the associated absorption systems, that differed from
those of the intervening systems, but concluded that differences in the results
among the three studies (Baker et al.  2002; Vestergaard 2003 and Ganguly et
al. 2001) could probably be attributed to various differences in selection of
the relatively small samples (~50-100 systems): among them, the most obvious
difference is the optical luminosity of the samples, progressively more
luminous in order of the references just cited.

Steep spectrum radio sources, from the above references, have an excess of
associated (\bet $<0.02$) absorbers and are thought to be viewed at a large
angle to the jet axis, so the sightline passes over the accretion disk (or
torus). Broad absorption line (BAL) systems are often thought to be viewed from
the same aspect, but are generally devoid of radio flux. There are suggestions
that the associated systems are related in some way to the broad absorption
line systems (Wampler et al. 1995; Baker et al. 2002; V03), summarized by V03.

There have also been  suggestions that narrow line QSO absorption line systems
(QSOALS) that have high velocities with respect to the QSO, i.e.  apparent
velocities of ejection up to tens of thousands of \kms $~$may also be
associated with the QSOs i.e.  they are close to the QSO ($<<$ 10 pc) but have
high relative velocity (Vanden Berk et al. 1996; Januzzi et al. 1996; Richards
et al. 1999; Richards 2001; V03; Misawa et al. 2007); candidate systems are
especially noticeable in the brightest QSOs and in radio sources with flat
radio spectra (often said to be core-dominated sources). Richards et al.
(2001), using the largest sample of radio QSOs available to date, showed that
there is an excess of C IV absorbers at high \bet $~$values in flat spectrum
radio sources (thought to be viewed close to the jet axis), strengthening
arguments of Richards et al. (1999) and Richards (2001) that up to 36\% of the
apparently high velocity material is associated with material intrinsic to the
background AGN. As shown by Misawa et al. (2007), such material may evidently
reach very high outflow velocities, up to and exceeding 20,000 \kms. Their
evidence for the intrinsic  nature of such components is the small filling
factor of the absorption lines in covering the background UV radiation source,
presumably the full accretion disk (Pereya et al. 2006). A primary question is
how to tell if a given system is intrinsic or not, whatever its \bet $~$value. 

In this paper, we use a large and homogeneous sample of associated
($z_{abs}\sim z_{em}$; relative velocity with respect to the QSO $<$ 3000 \kms)
{Mg~II} systems with $1.0<z_{abs}<1.86$ compiled from the Sloan Digital Sky
Survey (SDSS) Data Release 3 (DR3) catalog to determine the average dust
content, ionization and relative abundances in these absorbers. Our aim is to
determine if these systems are indeed intrinsic to the QSOs by (a) studying the
dependence of their average properties on QSO properties and (b) comparing the
properties of associated, {Mg~II} systems with those of intervening systems
(\bet $>$ 0.01) selected with similar criteria. In particular, we are looking
for a spectroscopic signature to distinguish associated from intervening
systems. 

We make use of the composite spectra of the sample (and various sub-samples
thereof), following the method recently advocated by York et al. (2006,
hereafter Y06).  In Section 2, we describe the criteria used for sample
selection, various sub-samples generated from the main sample and the method of
generating composite spectra. In section 3, we present our results which
include (i) a comparison between the properties of the samples of the
intervening and associated systems, using several statistical tests; (ii) a
discussion of the line strengths in the composite spectra of associated systems
compared to those for the intervening sample and of the state of ionization in
these systems; (iii) the measured extinction for various sub-samples; (iv) a
detailed analysis of the dependence of extinction and other properties on the
radio properties of the QSOs; (v) discussion of the abundances in associated
sample; and (vi) possible scenarios for the origin and location of the
absorbers.  A few systematic effects  that may be hidden in our data are noted
in section 4. Conclusions are presented in section 5. 
\section{Analysis}
\subsection{Sample selection}The absorption line system sample used here
was selected from the SDSS DR3 absorption line catalog compiled by York et al.
(2005, 2006).  In this section we describe the SDSS, the QSO sample in which
the absorbers are discovered, the construction of the SDSS absorber catalog,
and finally the selection of the associated absorption line systems used in
this study.
 
The SDSS (York et al. 2000, Stoughton et al. 2002)  is an imaging and
spectroscopic survey carried out from Apache Point Observatory, near Sunspot,
New Mexico, using a 2.5 meter telescope (Gunn et al. 2006). Candidate QSOs are
color-selected  from 5 color scans with a CCD camera (Gunn et al. 1998) of the
10,000 square degrees of the sky north of Galactic latitude 30 degrees. The
photometry from that imaging survey is based on a specially constructed set of
filters (Fukugita et al.  1996), calibrated with tertiary standards in the
imaging scans (Tucker et al. 2006). The establishment of secondary standards
using the SDSS Photometric Telescope (Hogg et al. 2001) and the 1-meter
telescope at the US Naval Observatory (Flagstaff), and their tie to a primary
standard, is described by Ivezic et al. (2004) and Smith et al. (2002). The
photometry is better than $\pm$ 0.02 magnitudes in all bands (Adelman-McCarthy
et al. 2007)

The QSOs are color selected (Richards et al. 2002a) from the reduced, archived
photometric data. Spectra are obtained using two, dual spectrographs. Plates
are drilled with 640, 3 arcsec diameter holes, distributed over an area of
seven square degrees, into which optical fibers are inserted by hand. About
100 of the fibers are allocated to QSOs. The selection of QSOs is meant to be
complete to magnitude 19.1 in the SDSS $i$ band, except that two objects within
55 arcsec of each other can not be observed using the same plate: one QSO must
sometimes be picked over another, or a galaxy may take higher priority over a
nearby QSO in the plate planning process. (Multiple, adjacent QSOs exist in the
archive because a second plate can be planned to overlap with a previous plate
(Blanton et al.  2003)). The astrometry, accurate to $<$ 0.1 arcsec,  that
allows precise placement of the fibers is described by Pier et al. (2003). The
plates are manually plugged and are used on nights not normally acceptable for
imaging.  Typically forty-five to sixty minutes of exposure are obtained in
several, 15-minute integrations, when cloudless skies prevail. The exposure
times or the total number of exposures are adjusted so the end result yields
(S/N)$^2\sim$7 for an object of $g$ magnitude 20.1. The selection of objects
yields a set of QSO spectra with about 70\% efficiency (some selected objects
are not QSOs), complete to better than 90\% (Vanden Berk et al. 2005). Each
plate includes fibers allocated to standard stars for flux calibration and for
sky removal.  Below $i$=19.1, QSOs are also, included in the selection if they
are X-ray sources from ROSAT (Voges et al. 1999) or radio sources from the
VLA FIRST (Faint Images of the Radio Sky at Twenty-cm) survey (Becker et al.
1995).

These spectra are reduced with two pipelines (two dimensional extraction and
one dimensional analysis). The QSO emission redshift is determined, by the
SDSS pipeline, by first assembling a list of emission peaks using a wavelet
based, peak-finding algorithm. These are matched against a standard set of
strong QSO emission lines to find the best-fit redshift (SubbaRao et al. 2002). 

The assembly of the archive of all QSOALSs in the spectra occurs in a separate
pipeline (York et al. 2005), run after the authoritative QSO spectra are
prepared for publication (Schneider et al. 2005 for DR3, Schneider et al. 2007
for DR5). A more complete description of the pipeline is in preparation (York
et al., in preparation).  Briefly, significant (equivalent width $>$ 3
$\sigma$, $\sigma$ being the uncertainty in the equivalent width
measurement due to noise in the spectrum; for narrow line QSOALS, the continuum
error is smaller, in general, than the error due to noise.), narrow (typically
3-8 pixels, each 70 \kms $~$in width) absorption features are identified after
fitting a smooth continuum to the spectrum. Poor night sky subtraction is
easily recognized and features found do not include such artifacts. BALs are
identified and flagged separately (Trump et al. 2006, for DR3, the data release
used in this paper): no lines are picked for this study from QSOs that contain
known BALs. However, most BALs in the SDSS are detected by the presence of C IV
absorption, and only a small fraction have detectable Mg II absorption (e.g.
Trump et al.  2006).  About half of the QSOs in our sample of associated
absorbers (described below) have \zem $<$ 1.5 and cannot be determined to be
free of BAL systems, since C IV emission is not redshifted into the SDSS
spectra until \zem $\ge$ 1.5. A search is first done for C IV doublets (one
line is picked and other line with the correct separation in wavelength is
searched for), then for {Mg~II} doublets, then for various Fe II pairs of lines
(in case one line of {Mg~II} is exactly at one of a select set of night sky
lines, or is obliterated by a poor correction for other lines).

Other lines are then selected by their correspondence to the {Mg~II}, Fe II or
C IV candidate systems. Unidentified lines are cataloged. These catalogs have
been carefully checked by hand to verify that our selection is not missing some
extreme situations that are not ``normal" but would be extremely interesting if
real\footnote{An example of a new type of system is given by D'Odorico (2007).
The lines of neutral Si I, Fe I and Ca I are seen for the first time in a
QSOALS. Our rules would still find this system, though the relatively weak
lines of Fe II and {Mg~II} and the even weaker lines of the noted neutral
species would only be visible in a few SDSS spectra with extremely high signal
to noise ratio.}. Thousands of systems have been examined and the rules above
are secure. Among lines identified as being within a line width of the main
lines noted above; those that are significant at the 4 $\sigma$ level; those
identified as being unambiguous (i.e., not possibly blended with a line from
another absorption line system); and those not in the Lyman alpha forest of
that QSO, are selected to provide a quality grade for each system. Four such
selected lines produces a grade A system; three lines, grade B system; two
lines, grade C system. Systems with only one line above a 4 $\sigma$
significance, and those with lines that that are all detected with a
significance between 3 and 4 $\sigma$, are, respectively, considered for
classification as grade D or E systems.  The lines used for grading are primary
lines usually seen in QSO absorption line systems: not all detected lines are
used, even if they fit all the other requirements.  For instance, Fe I is not
used, because any time that line would be seen, Fe II would be present and
stronger. This set of rules is constantly checked by eye, again, to not avoid
recognition of potentially interesting systems. The lines used for grading are
the strongest lines of Mg I, C II, {Mg~II}, Al II, Fe II, Al III, {C~IV} and Si
IV.  Under this set of protocols, for instance, C IV-only systems would be
given grades of C, while systems with significant C IV lines accompanied by a
line of Si IV or Mg II would be given grades of B, at least. If multiple Si IV
and/or Mg II, or gradable lines of other species are present in the systems,
the grades would be A.

The catalog thus constructed can be subjected to an SQL search that produces
the sample of systems that one is interested in. For this study, we selected
grade A systems, containing {Mg~II} lines with the stronger member of the
doublet having W$_{\rm Mg\;II}$ $>0.3$ {\AA} that are not in BAL QSOs; have
absorber redshifts 1.0$\le$ \zab $\le$1.86; have \bet $<$ 0.01; and are in QSOs
that have \zem $<$ 1.96. Each system was verified by visual inspection and any BALs present were eliminated. These systems form our main sample
(sample \# 1) consisting of 415 systems. The list of QSOs in this sample along
with their properties (\zem, \zab, \bet, \imag $~$and \dgi\footnote{This is
defined as the difference between the ($g-i$) color of a QSO and the median
value of ($g-i$) of all other verified SDSS QSOs with nearly the same redshift
(Richards et al.  2003) and it complements our use of the extinction curves to
derive $E(B-V)$.  ($E(B-V)\sim$\dgi /4 (Y06)).}) are given in Table 5, in the
appendix. The absorber rest-frame equivalent widths of the prominent lines,
along with 1 $\sigma$ errors, are given in Table 6 of the appendix.  About 100
out of these do have additional grade A and B systems (at absorption
redshifts $<<$ \zem) in their spectra. In principle, these can contribute to
the reddening of the parent QSO spectrum, but we show in the next section that
they do not contribute to the extinction. Lines in those systems do not
contribute to our composite spectrum used for measuring absorption lines
because they are masked before averaging the spectra.

The reasons for the particular selection of systems described above are as
follows. It is generally believed (e.g. Rao \& Turnshek 2000, Churchill
et al. 1999) that systems with \wmg $>$ 0.3 {\AA} have \nh $> 3\times 10^{17}$
cm$^{-2}$. Thus, the {Mg~II} line strengths were restricted so as to choose
systems that might have sufficiently large column densities in H I, so as to
have significant dust columns. The redshift range of the absorbers was chosen
to allow, on the one hand, only objects for which the 2175 {\AA} feature lies
completely within the SDSS spectrograph wavelength range, and, on the other
hand, to assure the {Mg~II} lines are at $\lambda<$8000 {\AA}, to avoid the
regions of the SDSS spectra that are contaminated by strong night sky emission
lines. The restriction on \zem was imposed to keep the Lyman $\alpha$
forest out of the SDSS spectra. We use the range $\beta<0.01$, to get the
purest sample of associated systems possible on the presumption that  the
higher the $\beta$ the more likely there are intervening systems mixed in. 
We wish to emphasize that the selection criteria were taken to
be same as those used by Y06 (except for the range of \bet $~$values).  This
was done so that we can compare the properties of the intervening systems
(studied by Y06) with those of associated systems (systems with \zab $\simeq$
\zem, studied here) without any selection bias. Such a comparison may indicate
some  signatures that may point towards the latter systems being intrinsic to
the QSOs. Our use of class A systems precludes including C IV-only or Mg
II-only systems, and we plan to study those systems in a later paper.

Having made the choice to study Mg II associated systems because we already
have a comparison sample of intervening systems, we nevertheless find that our
selection is typical, despite the preference of previous authors to focus
mainly on systems with C IV and higher ions. First, examination of the complete
set of systems selected (Table 5,6) reveals that 89\% have significant C IV
lines, when those lines are observable in SDSS spectra. Second, a preliminary
study of the statistics of all class A and B absorption line systems as a
function of $\beta$ (avoiding systems in BAL QSOs) revealed that the peak in
the number distribution of C IV systems for -0.003$<\beta<$0.01 is mimicked
when Mg II systems are used (without reference to whether they have C IV or
not) These statistics will be the subject of a future paper (Vanden Berk et
al., in preparation).

A remark is in order about the values of \bet $~$derived. Gaskel (1982) and
Wilkes (1984) showed that  the redshifts of QSOs differ from line to line.  As
summarized by Richards et al (2002b), the [O III] redshifts generally agree
with the stellar absorption line redshifts of the host galaxies of the QSOs,
when these can be observed. The shifts of {Mg~II} emission compared to the rest
frame narrow emission line regions defined by [O II] and [O III] is $<$ 200
\kms (Vanden Berk et al. 2001, Tytler \& Fan 1992) and probably about -100
\kms (Richards et al. 2002b). Richards et al. (2002b) and Richards (2006)
studied this effect in SDSS QSOs and advocated the removal of the systematic
shift mainly caused by the asymmetric profile of the C IV emission blend by
using the wavelength of 1546 {\AA} instead of the normal 1549 {\AA}.  This
procedure is adopted in the SDSS pipelines. For 250 QSOs in our sample, the C
IV line is available and used in deriving the redshift from the SDSS data, by
the SDSS pipeline. The species {Mg~II} and [C III] are available for all
spectra.  The system redshift reference, [O II] is covered in several spectra,
but not always detected. Figure 1 shows the comparison of relative
velocities of the absorption systems (with respect to the QSOs), namely
\bet$_{\rm SDSS}$, \bet$_{\rm [O~II]}$ and \bet$_{\rm Mg~II}$, respectively
obtained using the (a) SDSS emission redshifts (described above), (b) single
line [O II] redshfits when available (for 162 systems in our sample) and (c)
single line {Mg~II} redshifts (available for all QSOs in our sample, by
selection). The single line [O II] and Mg II redshifts are obtained from the
SDSS pipeline. It is apparent that the procedure adopted by the SDSS pipeline
does very well in picking the systemic redshifts of the QSOs, and we adopt them
in this paper.  These redshifts are statistically more precise than the single
line redshifts which are sometimes based on weak lines. The errors in \bet
$~$are generally $<$ 500 \kms $~$where we can check directly with [O II]. We
therefore feel the trends we discover later are not caused by errors in the QSO
redshifts used here.  Letawe et al. (2007), from an analysis of 5 objects,
have shown that the peak of H\bet $~$line may be a good indicator of systemic
redshifts.  However, for our sample (\zem $>$ 1) the H\bet $~$lines are outside
the SDSS spectra, and can not be used to determine the systemic redshifts. 
\subsection{Generation of arithmetic mean and geometric mean composite spectra}
Composite spectra in the absorber rest frame were generated in order to
compare continuum and absorption properties of various sub-samples of absorption
systems.  Both arithmetic mean and geometric mean spectra were generated; the
arithmetic mean produces a better representation of the average absorption
line profiles, while the geometric mean is better at preserving the continuum
properties of the QSOs.  The procedure for generating the composites is
summarized here, and fully described by Y06.

For the analysis of absorption lines, normalized arithmetic mean spectra were
generated as follows. The spectra of individual QSOs were corrected for
Galactic reddening (Fitzpatrick 1999; Schlegel et al. 1998) and (along with the
associated error arrays) normalized by reconstructions of the QSO continua,
using the first 30 QSO eigenspectra derived by Yip et al.  (2004). The
normalized spectra were shifted to the absorber rest frame and re-sampled onto
a common pixel-to-wavelength scale.  Pixels flagged by the spectroscopic
pipeline as possibly bad in some way (Stoughton et al. 2002) were masked and
not used in constructing the composites. Also masked were the pixels within 5
{\AA} of the expected line positions of detected absorption systems unrelated
to the target system. The normalized flux density in each remaining pixel was
weighted by the inverse of the associated variance, and the weighted arithmetic
mean of all contributing spectra was calculated for each pixel. The number of
individual spectra, and the distribution of absorber redshifts, contributing to
the composite spectrum of a particular line transition, varies from line to
line.  That is because the absorption system sample was selected to
simultaneously cover the Mg II doublet and 2175 {\AA} feature, which means that
other transitions may not be covered by all of the spectra. The rest wavelength
range of maximum sensitivity to absorption features in the composite spectra,
which gets contributions from all QSOs in the sample, is 1900 {\AA} to 3150
{\AA}: the rest frame spectrum of a QSO with absorption system at \zab $~$of 1
will cover the range 1900-4500 {\AA} while that of a QSO with an absorption
system at \zab $~$of 1.86 will cover the range 1330-3150 {\AA}. Thus, the C IV
lines are averages of spectra of QSOs with absorption systems with \zab $>$
1.4765, while the Ca II lines are averages of spectra of QSOs with absorption
systems with \zab $<$ 1.267. Accordingly, the apparent noise in the composite
spectra near these lines is greater than that near lines which fall in the rest
frame 1900-3150 {\AA} region (which gets contributions from all systems of the
sample). Therefore, the mean Ca II line and the mean C IV line cannot be
assumed to give an average picture of the relative behavior of the two lines. A
sub-sample must be confined to appropriate \zab $~$ranges so that all systems
averaged include coverage of all the lines to be compared.

For studying the dust content of the absorbers by its effect on the QSO
continua, geometric mean QSO spectra in the absorber rest frame were
generated.  The procedure for generating the geometric mean spectra is similar
to that for the arithmetic mean spectra, except that the individual spectra
were not normalized, and the arithmetic means of the logarithmic flux densities
of the non-masked pixels were calculated (producing geometric means in linear
flux density).  For each absorber sub-sample composite spectrum, a geometric
mean spectrum was generated for a QSO sample matched in SDSS $i$ magnitude and
QSO redshift, but without absorption line systems of grades A, B, or C in their
spectra. The matching of \zem $~$and \imag $~$means that the absolute $i$
magnitudes of the absorber and non-absorber pair also match. The matched,
non-absorber spectra were shifted in the procedure to the same rest frames as
their absorption sample counterparts, to produce composite non-absorber spectra
that can be compared directly to the composite absorber spectra.  There are
9737 QSOs in the SDSS DR3 data set available for use in the matched
non-absorber samples in the redshift range of interest here; they are free of
detected absorption systems, but otherwise satisfy the same selection criteria
as QSOs in the absorber sample. The details of the selection of the matching
non-absorber sample are described in Y06. The list of QSOs in the set of best
matched non-absorbers spectra (for the full absorber sample \#1) is given in
Table 5 of the appendix.  

The ratio of the geometric mean composite spectra of the absorber sample to
that of the non-absorber sample is the relative shape of the backlighting QSO
continuua, with and without absorption systems.  We interpret the ratio as the
extinction curve due to dust associated with the absorption systems.  It was
shown by Y06 that these ratios, produced using various sub-samples of
intervening absorption line systems, can be well fit by an SMC extinction law,
with different values of the color excess $E(B-V)$.  In the same way we fit SMC
extinction curves to the geometric mean spectra ratios for our current samples,
to derive values of $E(B-V)$ for associated systems.  In none of the absorber
samples did a Milky Way type extinction curve provide a better fit, and we
report only the results for the SMC type fits. 

To assess the uncertainty of the relative extinction curves, and the derived
values of $E(B-V)$, five independent (without any QSOs in common) matching
non-absorber samples were selected for each absorber sample\footnote{The sample
of Y06 was based on the SDSS DR1, for which there were fewer QSOs than we find
in DR3 and for which we could not find enough QSOs without absorption lines to
test the dispersion in color excess for the effects of random differences in
the QSO continua.}. The first non-absorber sample (given in Table 5 of the
appendix) selects the closest matching (in \zem $~$and \imag) non-absorber
QSOs. The subsequent non-absorber samples use the QSOs with next best matches,
not already used in the previous non-absorber samples. The largest average
differences between the \zem $~$and \imag $~$of the absorber and non-absorber
QSOs in the 5$^{th}$ non-absorber sample are $<\Delta z>$=0.03 and $<|\Delta
i|>$ =0.07.  Composite spectra were constructed and \ebv $~$values were
obtained for each of the five non-absorber samples for the absorber sample. The
five values of \ebv $~$were averaged to define the \ebv $~$of the absorber
sample. The rms dispersion of the five samples (of order $\pm$0.004 magnitudes,
attributable to sample selection randomness) was used as a measure of the
uncertainty in the values of $E(B-V)$.  The derived values of $E(B-V)$ and the
associated uncertainties are given in Table 1 for each absorber sub-sample.  We
conclude that for comparison of extinctions between our sub-samples,
differences of 0.01 in $E(B-V)$ are significant and are not caused by
systematic errors in defining the continuum.  It is possible that a few of the
non-absorber QSOs have high intrinsic extinction and the presence of such QSOs
in a non-absorber sample will reduce the value of \ebv. The use of five
independent non-absorber samples for determining the \ebv $~$values minimizes
the effect of such rare non-absorber QSOs. The sub-samples are described next
in \S\,2.3. 

\subsection{Sub-sample definitions} The full sample (\#1) was divided into
several sub-samples based on the absorber and QSO properties to study the
dependence of the dust content, ionization properties and relative abundances
on these properties.  The sub-samples were, initially, defined based on
dividing the full sample by absolute \imag\footnote{determined by using the
so-called ``concordance cosmology" ($\Omega_m=0.3$, $\Omega_l=0.7$, H$_0=70$
\kms Mpc$^{-1}$)}; {Mg~II} equivalent width; \bet $~$(positive or negative);
negative \bet $~$(further divided into two); and radio detection/non-detection
by the VLA FIRST survey. Specifically, we divided sample 1 into pairs of
sub-samples in three different ways: by absolute $i$ magnitude, at the median
value of -26.49 (\# 2 and 3); by W$_{\rm Mg\;II}$, at the median value of 1.35
{\AA} (\# 4 and 5); and by $\beta$, $<$ 0 or $>$ 0 (\# 6 and 7).  Sample 6 was
further subdivided into 2 sub-samples at the median \bet $~$value of -0.0022
(\# 8 and 9). 

In order to study the dependence of the properties of the associated absorbers
on the radio properties of the QSOs, the full sample was divided into
sub-samples based on the detection or non-detection of the QSOs by the FIRST
survey (for those QSOs covered by the FIRST survey); these sub-samples are
designated \#10 (hereafter, RD (radio-detected) QSOs) and \#11 (hereafter, RUD
(radio-undetected) QSOs).  The completeness limit of the FIRST survey catalog
across the survey area is 1 mJy (Becker et al. 1995); undetected QSOs have 20
cm fluxes less than this value. Miller et al (1990) define radio loud QSOs to
be those having 1.4 GHz luminosities greater than 10$^{25}$ W/Hz. Thus the RD
QSOs with \zem $<$ 2.0, in the FIRST survey are radio-loud. As all the QSOs in
our sample have \zem $<$ 1.96, our radio-detected and radio-undetected
sub-samples can be considered to be sub-samples of radio-loud and radio-quiet
QSOs according to the definition of Miller et al (1990).
 
In order to study the difference in ionization state (measured by the strength
of the {C~IV} absorption lines compared to the {Mg~II} absorption lines) of the
associated (\bet $<$ 0.01) and intervening (\bet $>$ 0.01) systems, we
constructed a sub-sample (\# 12) of associated systems with \zab $>1.4675$ so
that the wavelength of the C IV lines would be covered by the SDSS
spectrum\footnote{This sample is also definitely free of classic C IV, BAL
systems, which as previously noted, can not be assured for the lower $z$
systems.}. Sub-sample \#12 was sub-divided into two parts (\#13 and 14)
depending on \bet $~$being $>$ or $<$ 0. 

The required sub-samples of intervening systems, for comparison, are as
follows.  Y06 compiled a sample of 809 intervening systems (hereafter SY06)
satisfying the same criteria for selection as used here, except for having \bet
$>$0.01.  This forms our main sample of intervening absorbers. For ionization
studies, we selected a sub-sample of SY06 having \zab $>1.4675$ (sub-sample
SY06CIV). We also defined sub-samples based on radio detection or
non-radio detection, from the Y06 intervening sample, for comparison with the
associated sample. These two samples are referred to as SY06RD and SY06RUD,
respectively.   

Properties of various sub-samples (including those of intervening samples) are
listed in Table 1 which includes the defining criteria, the average and rms
dispersion of \ebv $~$values obtained by using the five independent,
non-absorber samples, along with the average values of W$_{\rm Mg\;II}$,
$\beta$, $z_{ab}$, $i$ magnitude, absolute $i$ magnitude calculated using the
concordance cosmology and \dgi $~$. 
\section{Results}
\subsection{Comparison of properties of intervening and associated systems} To
compare the properties of the associated (sample \#1) and the intervening
(SY06) systems, we have plotted in Figure 2 the distribution of W$_{\rm
Mg\;II}$, W$_{\rm Mg\;I}$/W$_{\rm Mg\;II}$, the doublet ratio of {Mg~II}, \walr
, $~$and absolute $i$ magnitudes for the two samples (left side of the Figures
2a to 2e).  The ionization measure, W$_{\rm C\;IV}$/W$_{\rm Mg\;II}$ for the
associated (\#12) and intervening (SY06CIV) sub-samples is shown in the left
side of Figure 2f (recall that both these sub-samples have spectra which cover
both C IV and {Mg~II} doublets).  Here, W$_{\rm Mg~I}$, W$_{\rm Al~II}$,
W$_{\rm Al~III}$ and W$_{\rm C\;IV}$ are the rest frame equivalent widths of
Mg~I $\lambda2852$, Al~II $\lambda1670$, Al~III $\lambda1854$ and C~IV
$\lambda1548$, respectively, in Angstroms. We use vacuum wavelengths throughout
and truncate the values in our ion notation. In making these plots we have used
only systems for which the equivalent widths of the lines are significant at
more than the 4 $\sigma$ level.  For sub-sample \#12, C IV lines were below 4
$\sigma$ for 26 systems, of the 250 total systems in the sub-sample. The upper
limits on W$_{\rm C\;IV}$/W$_{\rm Mg\;II}$ for these systems are $<$0.8. Thus
they will mostly fall within the first two bins.  Similarly, for the SY06CIV
sub-sample, the C IV lines were not detected at 4 $\sigma$ significance for 52
systems. The upper limits on W$_{\rm C\;IV}$/W$_{\rm Mg\;II}$ for these systems
are $<$0.5 and these systems will lie in the first two bins.  We have also
plotted in Figure 2, the distributions for associated systems with positive and
negative $\beta$ (sub-samples \#s 7 and 6 on the right of Figures 2a to 2e;
sub-samples \#s 13 and 14 on the right of Figure 2f).  

Qualitatively, it is clear from Figure 2 that, while the distribution for two
parameters: the Mg II doublet ratio and the absolute $i$ magnitude, are similar
between the associated and intervening samples, those for the three ion ratios
compared, and for the Mg II equivalent widths, are not. These conclusions also
seem to apply to the samples with positive and negative $\beta$ values. 

We have performed a number of statistical tests to determine the probabilities
that the quantities plotted in Figure 2, the equivalent width of C IV
$\lambda$1548, and the doublet ratio for {C~IV}, for the intervening and
associated systems, and also for sub-samples \#s 7 and 6, are drawn from the
same distributions. We have performed the KS test for three cases: (1) taking
only the measured values of various quantities and ignoring the non-detections,
(2) assuming the values for the non-detections to be zero, and (3) assuming the
values for the non-detections to be equal to 3 $\sigma$ values. The actual
result should be bracketed by these cases, in particular cases 2 and 3. In
addition, to directly account for the upper limits we have used two commonly
employed tests from survival analysis statistics, namely the Gehan and logrank
tests, which are described in astronomical nomenclature by (Feigelson \& Nelson
1985). Like the KS test, both survival analysis tests are designed to compare
the distributions of a parameter measured in two samples; unlike the KS test,
they take the upper limits within the distributions into account.  Both
survival analysis tests give similar results when applied to the current data
sets. The results of these tests are given in Table 2.  These show that the
doublet ratios of Mg II and C IV, and the absolute magnitudes of associated and
intervening systems are drawn from the same distributions, indicating a similar
degree of saturation of the Mg II and C IV lines for these two types of
systems. These conclusions are also valid for the two samples with positive and
negative $\beta$ values.  The equivalent width distribution for Mg II differs
for the comparison of associated to intervening samples, but not for the two
associated samples. The W$_{\rm C\;IV}$/W$_{\rm Mg\;II}$ ratios and possibly,
the W$_{\rm Mg\;I}$/W$_{\rm Mg\;II}$ ratios for the associated and intervening
systems are different, indicating a difference in ionization levels of the two
types of systems; the associated systems being more highly ionized as compared
to the intervening systems (see Figure 2). The situation is not so clear for
W$_{\rm Al\;II}$/W$_{\rm Al\;III}$ ratio, for which, the KS test and the
survival analysis tests give very different results. 

The situation regarding dependence of ionization of $\beta$ is less clear. The
results of KS tests for W$_{\rm C\;IV}$/W$_{\rm Mg\;II}$ and W$_{\rm
Mg\;I}$/W$_{\rm Mg\;II}$ do indicate a difference in ionization levels; the
negative $\beta$ systems being more highly ionized compared to the positive
$\beta$ systems (see Figure 2).  However, the survival analysis results do not
corroborate this. The distribution of W$_{\rm C\;IV}$ is however very different
for the two samples.

The ratio W$_{\rm Mg\;I}$/W$_{\rm Mg\;II}$ is enhanced among the associated
systems compared to the intervening systems, for ratios near 0.6 (SY06),
evidently mainly because of the contribution of systems with \bet$<$ 0, so the
statistical tests give low probability of the compared distributions being the
same.  This feature could represent an enhancement of {Mg~I} owing to higher
density in the negative \bet $~$systems, to a different temperature (affecting
the recombination rate of {Mg~I}; York \& Kinahan 1979) or to a specific effect
of the ionization field near 1100 {\AA} (the ionization limit of Mg I). (It was
found by Y06 that the blend of components that constitutes the Mg I feature
typically technically saturated at W$\sim$0.6 {\AA}, which could be related to
the excess seen here.)
\subsection{Line strengths in the composite spectra} Equivalent widths of the
measured lines in the arithmetic mean composite spectrum for the full sample
(\#1) are given in column 4 of Table 3.  Also given, in column 5, are the
equivalent widths of the lines for the sample of intervening systems (SY06) as
obtained by Y06. To the left are the vacuum wavelengths, the species and the
intrinsic strength indicator: $f \lambda^2/10^4$.  While the equivalent widths
of {Mg~II} $\lambda\lambda$2796,2890 and Al II $\lambda$1670 lines are similar,
the C IV $\lambda\lambda$1548,1550, {Si~IV} $\lambda\lambda$1396,1402 and Al
III $\lambda\lambda$1854,1862 (marginally) lines are stronger in the associated
sample than in the intervening sample. However, as noted above, because of the
selected range of $z_{em}$, the composite spectrum gets a contribution from all
absorbers only for lines with rest-frame wavelength in the range 1900-3150
{\AA}.  All the systems in sub-samples \#12 and SY06CIV, by construction,
contribute to the composite spectrum between wavelengths 1540 and 3150 {\AA}.
The equivalent widths of various lines for these two sub-samples are given in
columns 6 and 7 of Table 3. The arithmetic mean composite spectra of the two
sub-samples is shown in Figure 3. It can be seen that the higher ionization C
IV lines are indeed stronger and the lower ionization lines ({Mg~II}, Fe II, Al
II, Si II) are weaker in the associated systems as compared to those in the
intervening systems, consistent with the earlier conclusion that the ionization
is higher in associated systems. 

In order to determine if the ionization level in associated systems depends on
$\beta$, we measured the equivalent widths of lines in the composite spectra of
sub-samples \#13 and 14. These are also given in Table 3, columns 8 and 9,
respectively.  There is definite evidence of higher ionization for lower \bet
$~$values. The average absolute $i$ magnitudes of these two sub-samples
(-26.76 and -26.86  for \#13 and 14 respectively) differ only by 0.1 and the
distribution of absolute $i$ magnitudes is similar (KS test probability that
they are drawn from the same distribution being 0.13), so that the difference
in ionization is not caused by different intrinsic ionizing fluxes of the
QSOs. 

It has been found that, in a few QSOs, the absorbing material giving rise to
the associated systems covers the continuum source in the QSO only partially
(e.g. Barlow \& Sargent 1997; Hamann et al. 1997b). This can complicate the
interpretation of the observed equivalent widths. The partial coverage of the
source will change the doublet ratios of the absorption lines. The two
lines  of the C IV doublet are not completely resolved (detached) in our
composite spectra.  However, the distributions of the doublet ratios of
{Mg~II} lines for the intervening and associated samples, as well as the
sub-samples with different \bet $~$ranges are very similar (see Figure 2 and
Table 2). The equivalent widths and line profiles of {Mg~II} lines in sample
\#1 and SY06 are also very similar. So the issue of covering factor may not be
very important. Also, the equivalent widths of lines of C IV and {Mg~II} differ
in opposite senses in sub-samples with different \bet $~$values. Thus we
believe that our interpretation of higher ionization in associated absorbers as
compared to the intervening systems and its dependence on \bet $~$in associated
absorbers is justified.  

The strongest lines in Figure 3 undoubtedly consist of many components, some
saturated and some not. To understand the detailed behavior of the composite
spectra will require high resolution observations of a number of individual
systems, to fully sort out the effects of saturation, any dilution by low
covering factors that may exist in some cases and blending of many components.
Such a program has already been undertaken for the intervening sample (Meiring
et al. 2006; Peroux et al. 2006): differences between that sample and an
associated sample observed in the same way will be very useful for
understanding the associated systems. However the global ionization effects
noted here, from equivalent width ratios of the strong lines (of C IV and Mg
II), which show consistent behavior among various sub-samples, should not
change.

In Figure 4, we show profiles of a few selected lines in various sub-samples of
associated systems. It can be seen that most lines are stronger in the
sub-sample comprised of intrinsically faint QSOs (\#3) as compared to those in
the sub-sample comprised of intrinsically bright QSOs (\#2) (the average
observed $i$ magnitudes for the two sub-samples also differ considerably). This
was also observed by Y06 for the intervening systems and was understood as
being the effect of lower S/N ratio in the spectra of faint QSOs, which makes
only relatively stronger absorption systems in these QSOs detectable at the 4
$\sigma$ level. As expected, all the lines are stronger in the sub-sample of
stronger {Mg~II} systems (\#5) as compared to those in the sub-sample of weaker
{Mg~II} systems (\#4). As noted above, lines of higher ionization are stronger
in the sub-sample (\#6) of smaller \bet $~$compared to those in the sub-sample
(\#7) of larger \bet $~$values.  Finally, all of the lines are stronger in the
sub-sample (\#10) of RD QSOs as compared to those in the sub-sample (\#11) of
RUD QSOs. This appears to be due to the fact that sub-sample \#10 is comprised
of stronger {Mg~II} systems, having average W$_{\rm Mg\;II}$ = 1.9 {\AA}
compared to 1.52 {\AA} for sub-sample \#11. One of the systems in sub-sample
\#10 has very high \wmg ($\sim$ 9 {\AA}). This contributes significantly to the
equivalent widths of most lines in the composite spectrum of the sub-sample.  

The strong ionization effect noted in C IV is even more noticeable in Si IV
(Table 3, Figure 3 and Figure 4, row 4). This could be an effect of just a few,
peculiar systems with high enough redshift to contribute to both sample \#s 1
and 12. To check this, we formed a sample of systems with \zab $>$ 1.7378, so
that Si IV was completely covered in all cases, along with C IV and {Mg~II}.
The same trends evident in Table 3 persist. For respective sub-samples of (a)
intervening systems (from Y06); (b) associated, positive \bet $~$systems; and
(c) associated, negative \bet $~$systems, all  with \zab$>$1.7378, the values
of the equivalent widths of Si IV $\lambda$1393 are 380, 590 and 1330 m{\AA},
while for W$_{\rm C\;IV}$ we find 880, 1060 and 1320 m{\AA}, respectively.
These values are within 10\% of the corresponding number from sub-samples 12,
13 and 14 in Table 3, so the effect does seem to be real.  That is, the {Si~IV}
lines are relatively stronger as the systems have lower and lower $\beta$s.
The number of systems in these sub-samples are only 35, 27 and 45, so this
result needs to be confirmed in much larger samples (e.g., SDSS DR5).
\subsection{Extinction} The geometric mean composite spectrum for the full
sample (\#1) is given in Figure 5. Also plotted is the composite spectrum for
the matching non-absorber sample. In the bottom panel of the Figure we have
plotted the ratio of the two composites, the best-fit SMC extinction curve, and
the Milky Way extinction curve for the same value of $E(B-V)$. Similar to the
case of intervening systems (Y06), no 2175 {\AA} bump is seen in the observed
spectrum.  The bump is not present in the composite spectrum of any of the
sub-samples either, and the SMC extinction curve appears to describe the
observed extinction curve reasonably well. This is also supported by the fact
that the $E(B-V)$ values are close to 0.25 times the average values of \dgi
$~$for the sub-samples (see Table 1). Such a relationship was discovered by Y06
for the sub-samples of intervening systems, which also seemed to be well
described by an  SMC type of extinction curve. The lack of evidence for a
significant 2175 {\AA} bump is consistent with the fact that the feature has
been detected in the spectra of only a small number of individual QSOs (e.g.
Motta et al.  2002; Wang et al. 2004; Junkkarinen et al. 2004; Mediavilla et
al. 2005, see additional comments in Y06). 

By contrast to the case of intervening systems (Figure 2 of Y06 for instance),
the composite spectra for the associated systems (top panel of Figure 5 in this
paper) show typical QSO emission lines. This is due to the fact that for
associated systems, the absorption redshifts are very close to the emission
redshifts and the emission lines in individual spectra are very nearly aligned
even in the absorber rest-frames. In the extinction curves (in the bottom
panel), there appears to be some emission present at the wavelengths of [C III]
and {Mg~II} lines and possibly those of other QSO emission lines. This could be
due to either the emission by the absorbers, or an artificial effect produced
by the small differences in the emission redshifts of the absorber and
non-absorber QSOs. From Figure 5 one also gets the impression that the shape of
the emission lines in the absorber and non-absorber spectra are different. In
order to understand these effects we produced geometric mean composite spectra
of the absorber and non-absorber samples in the rest frame of the QSOs. The
ratio of these two composites did not show any emission. This shows that the
emission line profiles in the QSOs with associated absorbers and QSOs without
associated absorbers are, on average similar in shape.  This possibly also
indicates that the emission seen in the extinction curves of Figure 5 is not
real and is the effect of the difference in the emission redshifts of the
absorber and non-absorber QSOs. We can not however completely rule out emission
from the absorbers\footnote{The extended emission regions (EELR)  as seen in [O
II] and other lines discussed in the literature are mainly $<$ 30 kpc in size.
The three arcsec fibers of the SDSS spectroscopic survey include that region,
for all QSOs discussed here. Similar regions in radio galaxies show emission in
[C III] and {Mg~II} (McCarthy et al.  1993) which do not show up in the same
strengths in standard H II regions (Garnett et al. 1999). Hamann et al. (2001)
compute the emission line strengths of {Mg~II} EELR of 3C191, predicting an
equivalent width relative to the local QSO continuum of 4 {\AA} (1/3 of that
for [O II]). The combined equivalent width of the {Mg~II} absorption lines in
Figure 5 is 2.6 {\AA}, and the emission would be broader and more washed out.
Haiman \& Rees (2001) predict that infalling material could produce detectable
Lyman alpha emission from such core regions.  Evidently, detection of emission
from the EELR should be possible in SDSS composite spectra, if the intrinsic
QSO emission can be modeled precisely enough.}.
 
About 28\% of the QSOs in sample \#1 also have other, intervening, absorption
systems in their spectra. These, in principle, could contribute to the values
of \ebv $~$determined here. The effect is likely to be small in view of the
results of Y06, which showed that intervening systems with W$_{\rm Mg\;II}$
smaller than 1.53 {\AA} do not produce significant reddening and far outnumber
the systems with  stronger {Mg~II} absorption that do produce reddening.
However, to evaluate the effect of these intervening systems, we constructed
composites for the sub-sample of 298 associated systems (from our sample \#1)
which had no other absorption systems in their spectrum. The \ebv $~$for this
sub-sample is 0.029$\pm$0.003, which is the same as that of Sample \#1, to
within the errors.  We thus conclude that the other systems in 117 QSOs in our
sample \#1 are too weak to affect the value of \ebv $~$due to associated
systems determined here.

The extinction for the full sample (\#1; \ebv=0.026$\pm$0.004) is twice that
obtained for the intervening sample (SY06; \ebv=0.013), indicating a higher
amount of dust in the associated systems\footnote{This qualitative statement is
conservative.  That is, there is an evident discontinuity near 3650 {\AA} in
the composite spectra of the QSOs with {Mg~II} associated absorbers (top,
Figure 5). We assume that the true extinction is continuous and that the noted
discontinuity is a feature intrinsic to the QSO spectra or is an artifact of
the reductions.  This is reflected in our normalization at 3000 {\AA}, which
gives a lower limit to the color excess inferred.}. This could partly be due to
the following reason.  As noted in the last sub-section, the ionization level
is higher in the associated systems. Thus, by choosing systems with W$_{\rm
Mg\;II}$ $>0.3$ {\AA}, we have possibly chosen the associated systems with
higher total Mg column (that is, including all ionization states) and thus
higher total hydrogen column and therefore, higher dust column as compared to
those in the intervening systems.  We note that the average W$_{\rm Mg\;II}$ in
the intervening systems (1.73 {\AA}) is higher than that in the associated
systems (1.54 {\AA}).  This could be due to the ionization effect mentioned
above. It is however, not clear if the higher \ebv $~$in the associated systems
is due to this effect alone.  It is possible that the associated systems have
higher dust-to-gas ratio. It is also possible that part of the reddening is
caused by the dust intrinsic to the QSO, but not containing {Mg~II} (because of
ionization?). As shown in the next sub-section, the reddening is strongly
dependent on the radio properties of the QSOs. We return to
these points, below. 

The extinction in the sub-sample of faint QSOs (\#3) is somewhat higher
than that in the sub-sample of bright QSOs (\# 2). The average W$_{\rm
Mg\;II}$ for the fainter sub-sample is higher (1.70 {\AA}) compared to that for
the brighter sub-sample (1.37 {\AA}). This possibly indicates that the fainter
QSOs are fainter because of the higher dust content of the DLAs lying in front
of them. In the sub-sample of systems with W$_{\rm Mg\;II}>1.35$ {\AA} (\#5)
it is close to  two times that in the sub-sample of systems with W$_{\rm
Mg\;II}<1.35$ {\AA} (\#4).  Surprisingly, the value of \ebv $~$is 1.5 times
higher in associated systems with negative \bet $~$(\#6) than those with
positive \bet $~$(\#7).  Note that both these sub-samples have similar values
of W$_{\rm Mg\;II}$.  As noted in the previous sub-section, the ionization
seems to be higher in sub-sample \#6 than in sub-sample \#7.  The higher amount
of dust may thus indicate higher amount (by a factor $\sim$ 1.5) of the total
hydrogen (neutral plus ionized) in the latter. A similar dependence of \ebv
$~$on \bet $~$is observed if we divide the positive \bet $~$sub-sample (\# 7)
into two halves at the median value of 0.004. The lower \bet $~$sub-sample has
\ebv $~$which is higher than that of the high \bet $~$sub-sample. The effect is
also seen in sub-sample \#s 13 and 14. Thus the \bet $~$values seem to be an
indicator of the state of ionization of the absorbers and of the amount of
reddening.  Similar effects have been noted by Baker et al. (2002) and V03
using less quantitative means. The effect found here is much more subtle that
that claimed by Baker et al. (2002). The effect of radio properties of the QSOs
on the extinction is discussed in the next sub-section.  
\subsection{Dependence on radio properties of QSOs} In our sample of the
associated systems, 48 QSOs have non-zero FIRST flux (the RD sub-sample, \#10)
while 318 QSOs have non-detection in the FIRST survey (the RUD sub-sample,
\#11); the rest have not been observed by the FIRST survey. (Hereafter, we drop
reference to the FIRST survey, which is implied when we refer to radio
sources.)  In the SDSS DR3, the number of QSOs with \zem $~$between 1 and 1.96
is 23,914, in all of which we could have seen associated {Mg~II} absorbers if
present. Of these 1,728 are RD and 19,056 are RUD (the rest have not been
observed by the FIRST survey). As the selection criterion for the RD and RUD
sub-samples (as described in section 2.1) were identical and because the
average observed $i$ magnitudes (18.60 and 18.63 for the RD and RUD sub-samples
respectively), the average absolute $i$ magnitudes (-26.35 and -26.52 for the RD
and RUD sub-samples respectively) and \zab $~$(1.39 and 1.5 for the RD and RUD
sub-samples respectively) of the two sub-samples are almost equal, no other
biases are present and we can compare the frequency of occurrence of these
systems. Thus the incidence of associated absorbers with \wmg $>$ 0.3 {\AA} in
RD QSOs is 2.8\% while that in RUD QSOs is 1.7\%. RD QSOs are thus 1.7 times
more likely to have associated absorption as compared to the RUD QSOs.
Assuming binomial statistics, the probability of getting such a large
difference in the incidence of associated systems between the two sub-samples
is less than 1\%. 
\subsubsection{Intrinsic redness of radio-detected QSOs} As can be seen from
Table 1, the reddening in the RD QSOs with associated {Mg~II} systems is 
five times higher than that in the RUD QSOs. As noted in section 1,
associated C IV absorption may occur preferentially in steep-spectrum radio
sources. If, as is often assumed (see section 1), such sources are viewed in
the edge-on position, it is possible that the material in the accretion
disc/torus (unrelated to the absorption systems) may be causing part or even
most of the observed reddening in the radio-loud QSOs. The `redness' 
of radio QSOs has been noted earlier (e.g. Brotherton et al.  2001;
Baker et al. 2002; Ivezic et al. 2002).

To investigate this, we have, in Figure 6a, plotted histograms of \dgi $~$of
all RD (red lines) and RUD (blue lines) QSOs with $1.0<z_{em}<1.96$ (which is
the emission redshift range of our sample of associated absorbers) in the DR3
catalog (Schneider et al.  2005). It can be seen that the RD QSOs have a higher
fraction of red QSOs as compared to that in RUD sample (the red histogram is
shifted to the right). As many of the QSOs in the DR3 RD and RUD sub-samples may
have absorption systems, we have in Figure 6b, plotted similar histograms for
RD and RUD sub-samples of non-absorber DR3 QSOs having \zem $~$between 1 and
1.96.  There are 580 RD non-absorber QSOs (red lines) in this redshift range
while there are 6,786 RUD non-absorber QSOs (blue lines). According to Y06,
unreddened, unabsorbed QSOs have -0.2$<\Delta (g-i) <$ +0.2, consistent with
the RUD, non-absorber QSOs plotted in Figure 6b. The RD, non-absorber QSOs have
a much larger fraction of red QSOs as compared to the RUD non-absorber sample.
Thus, we have definite evidence of RD QSOs being intrinsically redder than the
RUD QSOs.

In order to quantify the intrinsic redness of the RD, non-absorber QSOs, we
constructed a composite spectrum of 250 RD, non-absorber QSOs in the QSO
rest-frame and compared it with similar composite spectrum of matching (in
\imag $~$and \zem) 250 RUD, non-absorber QSOs.  The KS test shows that the
\imag $~$and \zem $~$values in the two sub-samples are drawn from the same
distributions. The composite spectrum of the RD non-absorber QSOs is redder
than that of RUD non-absorber QSOs. An SMC extinction curve gives a good match
to the ratio of the two composites yielding a relative \ebv = 0.036$\pm0.0001$,
as the intrinsic average color excess in RD QSOs over RUD QSOs (see Table 4). 
\subsubsection{Dependence of extinction properties of associated absorbers on
the radio properties of QSOs} To test whether the higher \ebv $~$in RD QSOs is
solely because of the intrinsic dust in these QSOs or if the associated
absorbers in these QSOs also have higher dust content as compared to the rest
of the associated absorbers, we have in Figures 6a and 6b plotted the histograms of
\dgi $~$(black lines) for the sub-sample \#10. It is clear that the RD QSOs
with associated absorbers have a higher fraction of red QSOs as compared to the
RD non-absorbers (the black histograms extend more to the right).  Thus the RD
QSOs with associated absorbers are more likely to be reddened as compared to
those without such absorbers. While 56\% of RD QSOs with associated absorbers
have \dgi$>$0.2, the numbers for the RD and RUD non-absorber QSOs (with \zem
$~$between 1 and 1.96) are 26\% and 8\% respectively.  The three numbers for
\dgi$>$0.5 are 31\%, 9\% and 1.6\% respectively.  Evidently, studies of
extinction in radio QSOs must consider whether associated absorbers are present
or not, especially in individual cases. 

In order to quantify this result further, we compiled a matching sample of QSOs
(for the absorber QSOs in sub-sample \#10) from among the RD non-absorber QSOs.
The resulting \ebv $~$was found to be 0.062$\pm0.007$ (see footnote d in Table
1), smaller than the \ebv $~$of sub-sample \#10 in Table 1 by a factor of only
1.4, showing that though the RD, non-absorber QSOs are redder than the RUD,
non-absorber QSOs, a significant fraction of the reddening in our sub-sample of
RD QSOs is caused by the dust in the associated absorbers.

As noted in section 3.2, the average W$_{\rm Mg\;II}$ of the RD sub-sample
(\#10, 1.9 {\AA}) with associated {Mg~II} absorbers is higher than that of the
full sample (\#1, 1.54 {\AA}). Although, higher \wmg $~$is likely to be
indicative of a larger velocity dispersion of Mg II lines, a correlation
between \ebv $~$and \wmg $~$has been noticed by Y06 and is also seen in our
sample (see sub-sample \#s 4 and 5 in Table 1). Thus, the higher W$_{\rm
Mg\;II}$ of the RD sub-sample could be partially responsible for the higher
\ebv.  However, we note that the \ebv $~$for the RD sub-sample (\#10) is much
higher than for the sub-sample (\#5) of strong systems for which the average
W$_{\rm Mg\;II}$ is 2.23 {\AA} but \ebv $~$is only 0.034$\pm$0.003. Thus, the
somewhat higher average W$_{\rm Mg\;II}$ for the RD sub-sample compared to the
RUD sub-sample is not the sole reason for the high value of \ebv $~$which is
definitely related to the QSOs being RD.

If a large part of the excess extinction, observed in RD QSOs (over RUD QSOs),
is generated in the associated absorbers, then the \ebv $~$values in these QSOs
will depend on the absorber properties. In order to investigate this issue, we
divided the sub-sample of RD QSOs (\#10) into two parts, at the median value of
W$_{\rm Mg\;II}$ of 1.58 {\AA}. The relative \ebv $~$value of the high W$_{\rm
Mg\;II}$ sub-sample with respect to the low W$_{\rm Mg\;II}$ sub-samples
(obtained by fitting an SMC extinction curve to the ratio of the composite
spectra of the two sub-samples) is 0.092$\pm0.003$. Division into three equal
parts based on W$_{\rm Mg\;II}$ (W$_{\rm Mg\;II}<$ 1.22, 1.22$<$W$_{\rm
Mg\;II}<2.1$ and W$_{\rm Mg\;II}>$2.1 {\AA}) gives relative \ebv $~$(obtained
as explained above ) = 0.074$\pm0.001$ and 0.053$\pm0.001$ for the two pairs of
neighboring sub-samples (the higher W$_{\rm Mg\;II}$ sub-samples being redder).
The values of relative $E(B-V)$ are given in Table 4. It is clear that the
\ebv $~$is correlated with \wmg. The Spearman rank test to determine the
presence of a correlation between W$_{\rm Mg\;II}$ and \dgi $~$gives $R_s$ =
0.463 and the probability of chance correlation to be 0.001.  Thus it is very
clear that the higher reddening in the RD QSOs with associated absorbers is
strongly correlated with the strength of {Mg~II} absorption lines and should,
therefore, be due to the dust present in the absorbers.  The fact that RD QSOs
are intrinsically redder, evidently harboring more dust in their associated
systems (as compared to that in the systems associated with the RUD QSOs) may
indicate that the absorbing material is similar to that in the QSO and may thus
be intrinsic to the QSO. We do not, however, find any evidence of enhanced
abundances in the composite spectra (as in Figure 3) as discussed in section
3.5 below. 
\subsubsection{Comparison with the intervening sample} A significant
contribution to the \ebv $~$value for the full sample (\#1) must come from the
RD QSOs. For forty-nine objects in that sample, no radio observations are
available. Some of these could also have significant radio flux.  The RUD
sub-sample (\#11) has \ebv $~$of 0.016$\pm$0.004.  The QSOs in the matching
non-absorber samples for these do have some RD QSOs, so the value of \ebv=0.016
is a lower limit.  Restricting the non-absorber sample to RUD QSOs we get
\ebv=0.018$\pm$0.0007 (see Table 1 footnote e).  We have to compare these
values with \ebv $~$values for similarly selected sub-samples of the
intervening systems (SY06). Forty one systems in SY06 have been detected by the
FIRST survey while 614 are RUD; the remaining 129, have not been observed by the
FIRST survey. To avoid effects due to the selection of matching non-absorber
sub-samples, instead of determining the absolute \ebv $~$values for these
sub-samples, we directly determine relative \ebv $~$values between the RD and
RUD, intervening and associated sub-samples by fitting an SMC extinction curve
to the ratio of composite spectra of the sub-samples being compared.  We find
that the RD associated systems have a relative \ebv $~$of 0.062$\pm0.0001$ with
respect to the RD intervening systems. The RUD associated systems have a
relative \ebv $~$of 0.018$\pm0.001$ with respect to the RUD intervening
systems.  Thus the RD QSOs with associated absorbers are definitely more
reddened as compared to the RD QSOs with intervening absorbers. The RUD QSOs
with associated absorbers are also more reddened as compared to the RUD QSOs
with intervening absorbers. Thus on the whole, the associated absorbers are
dustier than the intervening absorbers.  Relative values of \ebv $~$ among
several sub-samples are given in Table 4.
\subsubsection{SDSS color-selected-radio-detected QSOs} As noted in section
2.1, SDSS QSOs are mostly color selected, but some QSOs (particularly those
below $i$ magnitude of 19.1) are selected because of their being ROSAT or FIRST
sources.  Such sources may be be redder than the average because they may not
satisfy the SDSS color-selection criteria, but instead have colors consistent
with the stellar locus. It is possible that the presence of such sources in the
RD sample (\#10) may be responsible for the excess reddening in this
sub-sample.  Ten sources in sub-sample \#10 are not color-selected while 2 of
the RD QSOs in SY06 are not color-selected. The relative \ebv $~$ of the
sub-sample of color-selected RD QSOs with associated systems with respect to
the color-selected RD QSOs with intervening systems (see Table 4) is
0.048$\pm0.0001$ (compared to 0.062 when the non-color selected QSOs are
included). Thus, the color-selected RD QSOs with associated absorbers are
significantly redder than the color-selected RD QSOs with intervening
absorbers.
\subsubsection{Radio Morphology} As noted above, the presence of associated
absorbers in radio QSOs may be correlated to the morphology and radio spectral
slope of these QSOs. Outflows from QSOs (e.g. jets or accretion disk winds) can
give rise to associated systems (e.g., Richards et al. 2001; Misawa et al.
2007). Models using beamed radio jet emission have been proposed to unify
flat-spectrum radio quasars and steep-spectrum radio galaxies (e.g., Orr \&
Browne 1982; Padovani \& Urry 1992). On average, a flat-spectrum quasar is
thought to be viewed closer to the jet axis than a steep-spectrum quasar. Here
we examine the radio morphologies and spectral indices  of quasars with and
without associated absorption with the hope of understanding the connection
between the direction of motion of the outflow (e.g. polar or equatorial) and
the absorber properties. 

To study  the radio morphology of quasars in our sub-samples, we proceeded as
follows: For each of the 48 RD quasars with associated absorbers (sample \#10),
we searched for sources detected in the FIRST image within 1.0' of the quasar's
optical (SDSS) coordinates. (At $1 < z < 2$, an angular separation of 1'
corresponds to about 500 kpc for the concordance cosmology.)  First, we
visually examined the morphologies in the FIRST images. If the source was
clearly resolved as a multi-component source (i.e., at least 1 distinct source
besides the core was found within about 30''), then we considered the source a
lobe (``L'') source.  If a single source, without a counterpart on the other
side, was found between 30''  and 60'' from the core, then it's likely not
related to  the quasar, and we treated the quasar as a core (``C'') source,
rather than a lobe source. For the objects with no other source found within
60'' of the core, we quantified the morphology using the procedures of Richards
et al.  (2001) and Ivezic et al.  (2002).  For each source, the FIRST catalog
lists the peak and integrated flux densities. Using these,  we determined a
dimensionless measure of concentration of radio emission, 
\begin{equation}
\theta = \large( {F_{\rm int} \over {F_{\rm peak}}} \large)^{1/2} 
\end{equation}
We classified the single sources with log $\theta^{2} \le 0.1$ as
core-dominated (``C'')  and those with log $\theta^{2} > 0.1 $ as partially
core-dominated (``PC'') sources.   (In a few ``C'' cases, the  values of log
$\theta^{2}$ are slightly negative due to the uncertainties in the FIRST flux
measurements.) Finally, we checked that  the above quantitative divisions
between ``C'' and ``PC'' are consistent with the contour maps of the FIRST
images.  (For only one ``C'' source in the absorber sample, large-scale
artifacts in the FIRST data give rise to log $\theta^{2} >  0.1$.) With these
definitions, the 48 quasars in sample \#10 consist of 26 ``C'', 8 ``PC'', and
14 ``L'' sources. Thus, $46 \%$ of the sources in the absorber sample are
spatially resolved (``PC'' or ``L''-type).

To understand how far these fractions relate to the existence of an associated
absorber along the sightline, we repeated the above procedure for the matching
48 RD quasars with no absorbers. Of these 48 quasars, 31 are ``C'' sources, 6
``PC'' sources, and 11 ``L'' sources.  Thus, 35 $\%$ of the sources in the
non-absorber sample are spatially resolved. There is thus a slight excess of
spatially resolved sources  in the absorber sample compared to the non-absorber
sample. Using binomial statistics, the probability of finding a fraction
of $46\%$ or more resolved sources in the absorber RD sample, if the true
fraction is only $35\%$, is about $0.08$.  Thus the difference between the
samples is not highly significant.  To confirm such an excess at the $1\%$
level would require a sample size of about 100 objects, which should be
possible to construct from the SDSS DR5 QSO data set.
 
We constructed absorber rest-frame composite spectra of the 26 ``C'' QSOs and of
the 14 ``L'' QSOs. As the line of sight to the ``L" QSOs is expected to
pass near or through the accretion disc, these QSOs are expected to be more
reddened. However, we find that the ``C'' composite is redder than the ``L''
composite, the relative \ebv $~$being 0.042.  This could very well be a small
sample effect and it will be very interesting to repeat this for bigger
samples.  

It is not possible to determine radio spectral indices for all of the quasars
in the RD absorber and non-absorber samples, since flux densities at
wavelengths other than 20 cm are not available for most of them. Of the 48
RD quasars with associated absorbers, only 15 have measurements at
other wavelengths [14 from the Green Bank (GB) 6 cm catalog (White \& Becker
1992), and one at 80 cm]. Using these along with the 20 cm FIRST measurements,
we calculated the spectral indices.  (In a few cases, the spectral index based
on the GB and FIRST measurements is somewhat different from that listed in
White \& Becker (1992), which probably results from  different resolutions of
the two surveys and possible variability between the two epochs of
observation.) In any case, taking the spectral index $\alpha_{\rm int}$ thus
calculated using integrated flux densities, we classified objects as
flat-spectrum (``F'') if $|\alpha_{\rm int}| \le 0.4$ and steep-spectrum
(``S'') otherwise. Of the 15 quasars in the absorber sample where we could
determine $\alpha_{\rm int}$, eight are ``F'' sources.  Of the seven
steep-spectrum sources, five have log $\theta^{2} \le 0.10 $ and are thus
likely to be compact. This suggests that these  associated absorbers are
associated with outflow material located close to the polar direction.  In the
non-absorber sample, 6 cm fluxes are available for seven quasars, of which four
are steep-spectrum sources (including one compact steep-spectrum source). Thus
compact steep-spectrum sources may be more common in cases with associated
absorbers. 

Overall, some fraction of the associated absorbers appear to arise in material
along the polar axis, while others arise in material away from the polar axis.
Larger samples are needed to study systematically correlations among the
absorber properties and outflow orientation and line-of-sight velocity.  
\subsection{Abundances} There  seems to be wide agreement that the abundances
in the broad emission line regions are super-solar (Ferland et al. 1996; Hamann
\& Ferland 1999; Hamann et al. 2002; Baldwin et al. 2003; Nagao et al. 2007;
Dhanda 2007) (and the same is true for the narrow line regions of Seyfert 2
galaxies; Groves et al. (2006)). There are a number of instances of suspected
super-solar abundances in associated absorbers (D'Odorico et al. 2004, Hamann
1997, Petitjean et al. 1994, Tripp et al. 1996). Fu \& Stockton (2006) discuss
the contamination of the broad line region and EELR by mergers with galaxies
with sub-solar abundances.  

It can be seen from Table 3 that for the weakest lines of dominant ions of  Mn
(3 lines near 2600 {\AA}), Fe ($\lambda\lambda$2260, 2249), Cr (3 lines near
2062 {\AA}) and Zn ($\lambda\lambda$2026, 2062), the ratios of lines of each
species indicate that the weakest lines are on the linear portion of the curve
of growth, independent of the Doppler widths or the width or saturation of the
stronger lines. We assume that Si II $\lambda$1808 is on the linear portion of
the curve of growth. (Note that the relative strengths of $\lambda$1808 and
$\lambda$1526 indicate that the Doppler width, in this case related to the
spread of components, is probably somewhat higher in sample \#1 than in sample
SY06, making our assumption conservative.) Then, the equivalent widths of the
weakest lines of these species are proportional to the column densities. We
find that the equivalent widths of the weakest lines of these ions are weaker
in sample \#1 (associated system average) than the lines of the intervening
averages (SY06): the ratios (sample \#1 divided by SY06) are 0.8, $<$ 0.8, 0.9,
1 and $<$ 0.8 for Si II, Cr II, Mn II, Fe II and Zn II, respectively. The value
for Zn II accounts for the fact that we can not deblend the Cr II 2062 line in
sample \#1 as we could in the SY06 sample. On the other hand, the same ratio
for Ni II might be as high as 5 for the weakest line, but of order 1.4 for two
other lines of comparable strength.

The extinction is twice as high in sample \#1 ($E(B-V)=$0.026) as in SY06
(0.013).  If the ratio of $E(B-V)$ to \nh $~$value is the same in associated and
intervening systems, then the abundances in sample \#1 are slightly less than in
SY06. There is no evidence that the associated systems containing Mg II have
greatly enhanced abundances that would indicate an association of the
associated systems with the gas from the QSO emission line region.

Of course, the dust might be different in the two types of systems, but the
extinction curves are of the same general shape. There could be ionization
corrections that would modify this indication about abundances: high
resolution observations and analysis with CLOUDY, as shown, for instance, by
Meiring et al. (2007), will be required to search for a stronger statement about
the abundances. Higher resolution is needed to resolve the individual
components in the systems and to resolve the contrary indication from the one
result for Ni II.

Thus, from our results there is no strong effect averaged over
415 associated absorbers studied here that would indicate a relation of those
absorbers to gas originating in the broad line regions. 

\subsection{Location and nature of the absorbers} We find several suggestions
of where the associated absorbers may be coming from. First, the ionization
effects in Si IV, C IV and {Mg~II} and their dependence on \bet $~$indicate
that lower \bet $~$systems are closer to the QSO. The various suggestions
referenced above that place some of the associated systems in the EELR, and the
confirmed existence of negative \bet $~$systems may be related to the large
widths (400 \kms) (Fu \& Stockton 2006) and bigger shifts (1800 \kms $~$to -600
\kms) of emission lines in the EELR (Christensen et al.  2006). While these do
not quite add up to negative $\beta$ values of –0.004 found here, the sample of
EELR is small: a larger sample may show that a wider range exists, comparable
to that seen in our {Mg~II} absorbers. An extensive study of the associated
systems found here, at high resolution, to determine abundances and populations
of fine structure excited states is in order.

We also note the anomaly in Mg I absorption line strengths mentioned earlier.
Hamann et al. (2001) found relatively strong Mg I lines in the associated gas
in 3C191 and draw an analogy to the strong Na I absorption in superwinds that
seem to have cool clumps within them (Heckman et al. 2000), that is most
evident when the galaxies are seen pole on and have moderately high (600-800
\kms) ejection velocities. Perhaps, with the same conditions but a luminous QSO
within tens of kpc, the temperature is tuned to allow a higher recombination
rate for Mg I owing to dielectronic recombination (York \& Kinahan 1979) and
the density is high enough to make Mg I particularly strong.  Hamann et al.
(2001) note that we have few spectroscopic diagnostics and few cases where we
can determine multiple diagnostics in associated systems: for example,
appropriate fine structure lines work only for gas densities of a few hundred,
for first ions but not for second ions; also, there are not many neutral
species detectable to establish photoionization rates, that might shed light on
the distances from the QSO. Likewise, a discrete and even higher density range
is accessible using the excited Fe II lines (Wampler et al. 1995). There may be
a wide range of conditions, but only a few that we can pick out and analyze
because of the unavoidable selection effects associated with atomic or ionic
parameters. Evidently, high resolution, high signal to noise observations may
reveal subtle examples of regions with a wider range of conditions, and provide
more insight into the likely origin of individual systems.

How can we tell if an apparently low $z$ (compared to \zem) absorber is intervening or is ejected
from the background QSO at high velocity?  Two attributes seem to distinguish
the associated absorbers (\bet $<$0.01)  studied here from the intervening
absorbers (\bet $>$0.01):  the higher ionization (specifically, the higher
ratio of W$_{\rm C\;IV}$ and probably W$_{\rm Si\;IV}$ to W$_{\rm Mg\;II}$, on
average), and the higher ratio of $E(B-V)$ to  W$_{\rm Mg\;II}$ in the
associated systems. The latter difference would be hard to distinguish in the
case of an individual absorber because of the range of intrinsic energy
distributions of QSOs at a given redshift (Vanden Berk et al. 2001, Richards et
al. 2002a, Richards 2006).  Whether the former can be a discriminate in
individual cases remains to be seen. In view of the much higher \ebv $~$values
for the associated systems in RD QSOs, we suggest that these are possibly
intrinsic to the QSOs.  Ganguly et al (2001) and V03 assumed that only C IV
systems (no low ions) were non-intervening systems. That appears from our study
not to be true.

A third possible distinguishing characteristic would be the  abundances, which
are possibly higher in associated systems in the sense of  being close to the
QSO, whatever the inferred velocity of ejection, if the gas comes from the
region of enhanced emission line abundances  of the AGN, as noted in references
at the end of section 3.5.  On the other hand, on average, there is little
evidence that the associated {Mg~II} absorbers have enhanced abundances.  In
light of the recent detection of intervening absorbers with super solar
abundances (e.g.  Khare et al. 2004; Peroux et al. 2006; Meiring et al. 2006),
this question cannot be answered  until there is a significant search for the
intervening galaxies near supposedly intervening QSO absorption line systems.
If  no galaxies are found near some high metallicity QSO absorption line
systems, the possibility of high velocity ejecta from the central region of the
AGN may need to be entertained.

The other well known indicators of intrinsic absorbers are the non-unity
covering factor and short time variability of the absorption lines\footnote{An
often mentioned criterion for a system being intrinsic is that it has variable
absorption. In light of the recent realization that a number of
``interstellar" lines in the Milky Way vary directly (Lauroesch \& Meyer, 2005;
Welty 2007) and that there are very small scale structures that could cause
variability owing to relative motion of the background source and subsequent
sampling of different parts of a foreground interstellar cloud, there is
evidently not yet one general technique that works in many individual cases.}.
The former does not appear to be widespread, as the effect does not appear in
our samples (but see Misawa et al. (2007) for a significant effect in weaker
lines at higher redshift than in our sample).  Apparently, only spectra of
individual objects, with high signal-to-noise, can be used to discern the
effect, which may be rare.  
\section{Selection effects in our study} The results presented here relating to
correlations with beta depend on the correctness of our values of $\beta$.
Richards et al. (2002b)  and Richards (2006) summarized many years of research
on the origin of the shifts between C IV and {Mg~II} emission lines in the same
QSO. They argue that the shifts arise from measuring the centroid of C IV
emission lines that are attenuated on the redward edge of the C IV profile,
shifting the centroid blueward. Further, the degree of this blueshift is more
extreme in sources that have bluer QSO continua, are not radio sources, and are
not X-ray sources. These same properties apply to the sources that have BAL
systems, that is, highly blueshifted absorption systems with much broader lines
than the $<$ 500 \kms $~$wide lines selected in our QSOALS catalog. Of course,
since the physics of the associated QSOALS and the BALs is equally uncertain,
there may be no sharp distinction between the origin of the two types of
absorption lines. At any rate, the origin and ionization of the associated
systems and the origin of the optical depth effects might be related, imposing
a correlation between $\beta$ and ionization from some source other than what
we discuss above.

We have showed that the redshifts derived using an empirical SDSS correction
procedure agree with the redshifts (hence, derived $\beta$ values for
absorbers) found from the narrow  [O II] lines, which should give the systemic
velocity of the QSO host galaxy. Thus, the effects nted above may not be very
important.  However, that line is not always available and we depend on
redshifts from broad lines in particular cases.

Other subtle correlations between absorption lines and the background QSOs
might affect the present study. (a) The negative velocity and positive
velocity, apparent motions of the associated absorbers may not be physically
related in a continuous fashion: the negative and positive $\beta$ systems
might have a different physical origin. (b) With larger samples and further
sharpening of the definitions of ``associated", ``intervening" and ``ejected",
it may be possible to further refine the selection of ``associated" systems. It
may be that in our sample we have mixed in some true intervening systems that
are, nonetheless, near the QSO, or that some ejected systems, destined for
higher velocities than 3000 km/sec, have been included in our ``associated"
sample. Richards et al. (1999) suggest that one may have to pick a pure
intervening sample from systems that are have $\beta >$ 0.1 to be free of
ejected material. (c) Our use of the absolute $i$ magnitude to construct
matched pairs may not completely control for intrinsic variations in the QSO
continuum among absorber and non-absorber samples. X-ray, radio and infrared
luminosities may be needed to be matched for a thorough removal of such
effects.  Measurements of these are often unavailable, or are only available to
flux limits of surveys that could be inadequate for the task.  (The best
available X-ray and near-IR survey data matched to the SDSS data, are described
by Anderson et al. 2007 and Ivezic et al. 2002). 

Finally, we caution that the trends in the state of ionization of the
absorbers, as noted by us, need to be confirmed by higher resolution
observations as line saturation is likely to be significant for Mg II lines.
\section{Conclusions}
We have studied a sample of 415 intermediate redshift, {Mg~II} associated
systems in the spectra of SDSS QSOs. Our main conclusions are as follows.
\begin{enumerate}
\item There is definite evidence of dust extinction in the associated systems.
The average extinction is two times that found in a sample of similarly
selected, intervening systems. This larger extinction could be attributed to
higher average N$_{\rm H\;I}$ +N$_{\rm H\;II}$ in associated compared to
intervening systems, or to differences in the dust to gas ratio.
\item There is no evidence for the 2175 {\AA} bump in the extinction curves of
the associated absorbers.
\item The extinction curve for these absorbers is similar to that of the SMC,
which has also been found to be the case for intervening absorbers.
\item Associated absorbers are in a higher state of ionization compared to the
intervening absorbers.
\item In the relative velocity range 0.01$>$\bet$>$-0.004 studied, the
ionization conditions and the total extinction in the associated systems is a
function of their apparent relative velocity with respect to the QSO, systems
with lower relative velocity being more ionized and more highly reddened.
\item There is no obvious evidence for higher abundances in the associated
systems than those in the intervening systems.
\item About a third of the absorption systems do have redshifts higher than the
emission redshift of QSOs and thus appear to be infalling. The direct inference
would be that lower \bet $~$systems are closer to the ionization sources. No
clues are found that distinguish the material as ambient material (shocked?) in
the QSO host galaxies or as returning gas from QSO jet or accretion disk
outflows.
\item The QSOs with non-zero radio flux in the FIRST survey are intrinsically
redder than QSOs with null detection in the FIRST survey by relative \ebv
$\sim$0.04, on average, for an SMC extinction curve, even when there are no
absorption lines present.
\item Associated systems in QSOs with non-zero radio flux in the FIRST survey
have three to four times as much dust as those in QSOs with null detection in
the FIRST survey. This excess reddening is correlated to the strength of Mg II
absorption lines and appears to originate in the absorbers themselves. These
absorbers are thus, significantly different from the intervening systems and
may possibly be intrinsic to the QSOs.
\item No clear discriminant between associated and  intervening systems has
been found that would definitely work on a  case by case basis, despite the
clear but subtle differences in ionization and reddening noted above.
``Intrinsic" but high ejection velocity systems may be hard to discern among the
intervening systems, except through non-detection of a galaxy  at the absorber
redshift.
\end{enumerate}
\appendix
\section{The absorber and matching non-absorber samples} The list of absorbers
in the full absorber sample (\#1) and the corresponding, best matching
non-absorber sample is given in Table 5. The Table lists plate, fiber and
MJD numbers, \zem, \dgi $~$and \imag $~$for the absorber and the non-absorber
QSOs. Also listed are the \zab and \bet $~$values for the absorber sample. In
Table 6 we give the absorber rest-frame equivalent widths, and their 1 $\sigma$
errors, of Al II $\lambda$ 1670, C IV $\lambda\lambda$ 1548,1550, Mg I
$\lambda$ 2852, Mg II $\lambda\lambda$ 2796, 2803, Si IV $\lambda\lambda$
1396,1403 and Fe II $\lambda$ 2382 for the absorption systems. The equivalent
widths are in {\AA}.  Here, we only give small parts of these tables.  The full
tables can be accessed electronically. 
\acknowledgements
DVB and DPS acknowledge the support from the NSF grant AST06-07634. PK
acknowledges support from the Department of Science and Technology (Govt. of
India) grant SP/S2/HEP-07/03. DGY is grateful for support of students by the
Kavli Institute. GTR was supported in part by a Gordon and Betty Moore
Fellowship in data intensive sciences at JHU. VPK acknowledges support from the
NSF grant AST 06-07739 to the University of South Carolina.  

Funding for the SDSS and SDSS-II has been provided by the Alfred P.
Sloan Foundation, the Participating Institutions, the National
Science Foundation, the U.S. Department of Energy, the National
Aeronautics and Space Administration, the Japanese Monbukagakusho,
the Max Planck Society, and the Higher Education Funding Council for
England. The SDSS Web Site is http://www.sdss.org/.

The SDSS is managed by the Astrophysical Research Consortium for the
Participating Institutions. The Participating Institutions are the
American Museum of Natural History, Astrophysical Institute Potsdam,
University of Basel, University of Cambridge, Case Western Reserve
University, University of Chicago, Drexel University, Fermilab, the
Institute for Advanced Study, the Japan Participation Group, Johns
Hopkins University, the Joint Institute for Nuclear Astrophysics, the
Kavli Institute for Particle Astrophysics and Cosmology, the Korean
Scientist Group, the Chinese Academy of Sciences (LAMOST), Los Alamos
National Laboratory, the Max-Planck-Institute for Astronomy (MPIA),
the Max-Planck-Institute for Astrophysics (MPA), New Mexico State
University, Ohio State University, University of Pittsburgh,
University of Portsmouth, Princeton University, the United States
Naval Observatory, and the University of Washington.

\clearpage
\begin{table}
{\scriptsize
\caption{Sample definitions and properties}
\vspace*{0.1in}
\begin{tabular}{|l|l|l|r|l|l|r|l|l|l|}
\hline
\multicolumn{1}{|l|}{\bf Sample}&\multicolumn{1}{|l|}{\bf Selection
}&\multicolumn{1}{|l|}{\bf E(B-V)}&\multicolumn{1}{|l|}{\bf Number}
&\multicolumn{1}{|l|}{\bf $<$\wmg$>$}&\multicolumn{1}{|l|}{\bf
$<z_{abs}>$}&\multicolumn{1}{|l|}{\bf $<\beta>$}&\multicolumn{1}{|l|}{\bf
$<i\tablenotemark{a}>$}&\multicolumn{1}{|l|}{\bf$<$M$_i\tablenotemark{b}>$}&\multicolumn{1}{|l|}{\bf
$\Delta(g-i)$\tablenotemark{c}}\\ 
{\bf Number}&{\bf criterion
}&&\multicolumn{1}{|c|}{\bf of systems}&\multicolumn{1}{|c|}{\bf in {\AA}}&&&&&\\
\hline
1&Full sample&0.026$\pm$0.004&415&1.54&1.48&0.0018&18.61&-26.52&0.124\\
2&M$_{i < -26.49}$&0.024$\pm$0.008&208&1.37&1.62&0.0019&18.31&-27.08&0.082\\
3&M$_{i\ge -26.49}$&0.034$\pm$0.006&207&1.70&1.35&0.0019&18.92&-25.95&0.166\\
4&W$_{\rm Mg\;II}<$1.35 {\AA}&0.018$\pm$0.005&208&0.86&1.51&0.0017&18.49&-26.70&0.087\\
5&W$_{\rm Mg\;II}\ge$1.35 {\AA}&0.034$\pm$0.003&207&2.23&1.46&0.0020&18.74&-26.34&0.161\\
6&\bet$ <$ 0.0&0.035$\pm$0.004&152&1.54&1.49&-0.0027&18.58&-26.54&0.140\\
7&0.0$\le$\bet $<$ 0.01&0.020$\pm$0.005&263&1.54&1.48&0.0045&18.63&-26.51&0.115\\
8&\bet$ <$ -0.0022&0.033$\pm$0.006&74&1.65&1.57&-0.0045&18.45&-26.82&0.132\\
9&$-0.0022\le\beta <$ 0.0&0.038$\pm$0.006&78&1.43&1.41&-0.0011&18.70&-26.27&0.148\\
10&Radio-detected (RD)&0.086$\pm$0.003\tablenotemark{d}&48&1.90&1.39&0.0012&18.60&-26.35&0.359\\
11&Radio-undetected (RUD)&0.016$\pm$0.004\tablenotemark{e}&318&1.52&1.50&0.0019&18.63&-26.52&0.094\\
12&\zab$>$1.4675&0.025$\pm$0.003&250&1.47&1.65&0.0019&18.64&-26.80&0.114\\
&\bet$<0.01$&&&&&&&&\\
13&\zab$>$1.4675&0.021$\pm$0.005&156&1.48&1.65&0.0049&18.69&-26.76&0.114\\
& positive \bet&&&&&&&&\\
14&\zab$>$1.4675&0.031$\pm$0.005&94&1.46&1.65&-0.0032&18.57&-26.86&0.115\\
& negative \bet&&&&&&&&\\
SY06&Sample 1 &0.013&809&1.73&1.33&0.120&18.96&-26.43&0.042\\
&from Y06\tablenotemark{f}&&&&&&&&\\
SY06CIV&SY06 with&0.015&250&1.69&1.61&0.064&18.99&-26.66&0.051\\
&\zab$>$1.4675&&&&&&&&\\
SY06RD&Radio-detected&0.037&41&1.98&1.31&0.11&19.09&-26.29&0.16\\
& from SY06&&&&&&&&\\
SY06RUD&Radio-undetected&0.015&614&1.73&1.34&0.12&19.05&-26.42&0.035\\
& from SY06&&&&&&&&\\
\hline
\end{tabular}
	
\tablenotetext{a}{SDSS $i$ magnitude, corrected
for Galactic extinction}
\tablenotetext{b}{Absolute $i$ magnitude, using the concordance cosmology}
\tablenotetext{c}{$\Delta(g-i)$ for absorber sample ($\sim
4E(B-V)$, see text)}
\tablenotetext{d}{Restricting the matched non-absorber comparison sample to
radio-detected QSOs reduces the value of \ebv $~$to 0.062$\pm$0.007.}
\tablenotetext{e}{Restricting the matched non-absorber comparison sample to
radio-undetected QSOs increases the value of \ebv $~$to 0.018$\pm$0.007.}
\tablenotetext{f}{This sample uses the same selection criterion
as the sample \#1 of associated systems but has \bet$>0.01$}
}
\end{table}
\begin{table}[h]
\caption{Results of statistical tests\tablenotemark{a}}
\vspace{0.1cm}
\begin{tabular}{|l|l|l|l|l|l|l|}
\hline
\multicolumn{1}{|c|}{\bf Variable}&\multicolumn{1}{|c|}{\bf
Sample}&\multicolumn{1}{|c|}{\bf KS}&\multicolumn{1}{|c|}{\bf Survival}&\multicolumn{1}{|c|}{\bf Sample
}&\multicolumn{1}{|c|}{\bf KS}&\multicolumn{1}{|c|}{\bf Survival}\\
&\multicolumn{1}{|c|}{\bf \#s}&\multicolumn{1}{|c|}{\bf
Test\tablenotemark{b}}&\multicolumn{1}{|c|}{\bf Analysis\tablenotemark{c}}&\multicolumn{1}{|c|}{\bf \#s }&\multicolumn{1}{|c|}{\bf
Test\tablenotemark{d}}&\multicolumn{1}{|c|}{\bf Analysis\tablenotemark{e}}\\
\hline
W$_{\rm Mg\;II}$ & 1, SY06  & 0.00 &0.00, 0.00& ~6, 7 & 0.43&0.46, 0.51\\ 
{Mg~II} doublet& 1, SY06  &   0.12  &0.26, 0.39& ~6, 7&0.53&0.78,  0.68\\
~~~~~~~~~~~~ratio&&&&&&\\
W$_{\rm Mg\;I}$/W$_{\rm Mg\;II}$ & 1, SY06    & 0.01, 0.00, 0.36 & 0.01, 0.05&~6, 7 &0.03, 0.00, 0.12&0.26, 0.11\\
W$_{\rm Al\;II}$/W$_{\rm Al\;III}$ & 1, SY06   &0.10, 0.00, 0.00  & 0.70, 0.62&~6, 7 &0.45, 0.34, 0.55&0.76, 0.93\\
W$_{\rm C\;IV}$/W$_{\rm Mg\;II}$  & 12, SY06CIV &   0.00, 0.00, 0.00 & 0.00, 0.00& 13, 14  &0.04, 0.05, 0.05&0.13, 0.39\\
W$_{\rm C\;IV}$ & 12, SY06CIV  &0.05, 0.00, 0.00   & 0.12, 0.08 &~6, 7 &0.08, 0.09, 0.15 &0.06, 0.07\\
C IV doublet& 12, SY06CIV  & 0.77,  0.68, 0.58 &0.15, 0.14 &~6, 7 &0.62, 0.92, 0.82&0.25, 0.42\\
~~~~~~~~~~~~ratio&&&&&&\\
Absolute $i$& 1, SY06  &0.13&0.21, 0.25&~6, 7&0.96&0.07, 0.10 \\
magnitude&&&&&&\\
\hline
\end{tabular}
\tablenotetext{a}{Probability values smaller than 5$\times 10^{-3}$ have been
given as 0}
\tablenotetext{b}{Probability that the associated and intervening samples (left
panels of Figure 2, a-f and the equivalent width of C IV $\lambda1548$) are
drawn from the same distribution. Three numbers in this column (which should
bracket the actual result) indicate KS test probabilities for the cases (1)
non-detections have been excluded, (2) non-detections are treated as having
zero equivalent width, and (3) non-detections are are replaced by 3 $\sigma$
values, respectively; a single entry is given for the quantities whose values
are available for all the systems.}
\tablenotetext{c}{Probability that the associated and intervening samples (left
panels of Figure 2, a-f and the equivalent width of C IV $\lambda1548$) are
drawn from the same distribution. The two numbers are for Gehan test and the
logrank test, respectively.  These tests include the 3 $\sigma$ upper limits.}
\tablenotetext{d}{Probability for samples with positive and negative $\beta$
(right panels of Figure 2, a-f and the equivalent width of C IV
$\lambda1548$) to be drawn from the same distribution. The three numbers are
for cases described in footnote b.}
\tablenotetext{e}{Probability for samples with positive and negative $\beta$
(right panels of Figure 2, a-f and the equivalent width of C IV $\lambda1548$)
to be drawn from the same distribution. The two numbers are for Gehan test and
the logrank test, respectively.  These tests include the 3 $\sigma$ upper
limits.}
\end{table}
%
\begin{table}
{\scriptsize
\caption{Equivalent widths for selected samples}
\vspace{0.1in}
\begin{tabular}
{|l|l|r|r|r|r|r|r|r|}
\cline{1-9} 
\multicolumn{3}{|c|}{Sub-sample number}&1&SY06 &12 &SY06CIV&13 &14  \\
\hline
\multicolumn{3}{|c|}{$\beta$}&$<$0.01 &$>$0.01 &$<$0.01 &$>$0.01 &0.0-0.01 & $<$0 \\
\hline
\multicolumn{3}{|c|}{Number of systems}& 415& 809& 250& 250&156 &94  \\
\hline
$\lambda $& Species& ${f\lambda^2\over{10^4}}$\tablenotemark{a}&\multicolumn{6}{|c|}{Equivalent
width with 1 $\sigma$ errors, in m{\AA}; upper limits are 3 $\sigma$ values}\\
\cline{1-9} 
1548.20&C IV&45.5&1146$\pm$8&839$\pm$10&1148$\pm$8&855$\pm$8&1131$\pm$13&1216$\pm$22 
\\
1550.78& C IV&22.8&880$\pm$8&580$\pm$9&881$\pm$8&625$\pm$7&738$\pm$13&1082$\pm$23 
\\
2852.96& Mg I&1489&252$\pm$4&282$\pm$4&213$\pm$6&261$\pm$7&201$\pm$7&224$\pm$11 
\\
2796.35& {Mg~II}&482&1388$\pm$6&1432$\pm$4&1167$\pm$8&1465$\pm$9&1228$\pm$10&1112$\pm$12
\\
2803.53& {Mg~II}&241&1248$\pm$6&1261$\pm$4&1102$\pm$8&1255$\pm$9&1134$\pm$10&1075$\pm$12
\\
1670.79& Al II&486&464$\pm$6&471$\pm$6&431$\pm$8&455$\pm$6&430$\pm$8&442$\pm$10
\\
1854.72& Al III&192 &229$\pm$4&183$\pm$4&240$\pm$6&201$\pm$7&208$\pm$8&269$\pm$9
\\
1862.79& Al III& 97&128$\pm$7&112$\pm$4&131$\pm$6&111$\pm$6&112$\pm$9&132$\pm$10
\\
1808.00& Si II&0.69 &53$\pm$5&68$\pm$5&52$\pm$5&57$\pm$4&46$\pm$7&61$\pm$8
\\
1393.32& Si IV&99.7&890$\pm$28&\tablenotemark{b}&900$\pm$26&380$\pm$19&581$\pm$23&1423$\pm$45
\\
1402.77& Si IV&50&605$\pm$20&\tablenotemark{b}&608$\pm$19&324$\pm$15&300$\pm$23&1100$\pm$46
\\
1526.71& Si II&31&473$\pm$7&383$\pm$10&471$\pm$7&397$\pm$9&404$\pm$10&556$\pm$11
\\
2056.26& Cr II&43.6&22$\pm$4&39$\pm$4&23$\pm4$&23$\pm$4&26$\pm$8&23$\pm$8
\\
2062.24&Cr
II&32.4&19$\pm$3\tablenotemark{c}&27&19$\pm4$\tablenotemark{c}&40$\pm$3\tablenotemark{c}
&33$\pm8\tablenotemark{c}$&21$\pm$7\tablenotemark{c}\\
2066.16& Cr II&21.7&$<11$&14$\pm$4&14$\pm4$&17$\pm$3&$<18$&$<24$
\\
2576.88& Mn II&240&52$\pm$4&54$\pm$4&51$\pm$4&64$\pm$6&42$\pm$6&74$\pm$8
\\
2594.50& Mn II&188&33$\pm$3&50$\pm$4&17$\pm$4&44$\pm$8&25$\pm6$&$<15$
\\
2606.46& Mn II&135&23$\pm$2&26$\pm$4&9$\pm$3&24$\pm$7&27$\pm$6&10$\pm$3
\\
2382.77& Fe II&182&760$\pm$8&761$\pm$4&674$\pm$8&672$\pm$5&607$\pm$11&761$\pm$8
\\
2600.17& Fe II&162&709$\pm$3&783$\pm$4&597$\pm$4&736$\pm$8&667$\pm$6&517$\pm$7
\\
2344.21& Fe II&62.6&486$\pm$4&550$\pm$4&413$\pm$7&457$\pm$4&415$\pm$7&424$\pm$11
\\
2586.65& Fe II&46&428$\pm$3&509$\pm$4&343$\pm$4&475$\pm$8&382$\pm$6&297$\pm$6
\\
2374.46& Fe II&17.6&259$\pm$5&298$\pm$4&214$\pm$6&257$\pm$6&220$\pm$9&213$\pm$10
\\ 
1608.45& Fe II&14.9&$128\pm$5&$\pm$&121$\pm$5&210$\pm$7&131$\pm$6&111$\pm$8
\\ 
2260.78& Fe II&1.2&43$\pm$4&46$\pm$4&34$\pm$5&34$\pm7$&25$\pm7$&51$\pm$11
\\
2249.88& Fe II&0.9&35$\pm$5&35$\pm$4&19$\pm$4&13$\pm$4&$<18$&34$\pm$9
\\
2367.59& Fe II&0.012&$<7$&$<$8&$<9$&$<13$&$<16$&$<21$
\\ 
2012.17& Co II&15&12$\pm$4&15$\pm$4&$<16$&12$\pm3$&$<25$&$<18$
\\
1941.29& Co II& 12.8&$<8$&$<$8&$<12$&$<10$&$<12$&$<20$
\\
1741.55& Ni II&13 &33$\pm$4&28$\pm$4&32$\pm5$&$<24$&34$\pm$8&28$\pm$6
\\
1709.60& Ni II&9.4 &22$\pm$4&16$\pm$4&19$\pm$3&12$\pm3$&13$\pm$3&$<13$
\\
1751.92& Ni II&8.6 &26$\pm$4&$<6$&21$\pm4$&10$\pm3$&33$\pm$8&$<9$
\\
2026.14&Zn II\tablenotemark{c}&205&46$\pm$5&48$\pm4$&44$\pm$6&36$\pm$3&30$\pm$5&59$\pm$11
\\
2062.66&Zn
II&105&19$\pm$3\tablenotemark{c}&22&19$\pm4$\tablenotemark{c}&40$\pm$3\tablenotemark{c}&33$\pm8$\tablenotemark{c}&21$\pm$6\tablenotemark{c}
\\
\cline{1-9} 
\end{tabular}
}
\tablenotetext{a}{This is the relative strength for lines of the same species
and can be used to indicate the degree of saturation present.  For instance,
for sample 1, W$_{\rm Al\;II}$/W$_{\rm Al\;III}$ = 1.8, whereas the ratio in
column 3 is 2: on average, the doublet lines of Al III are only slightly
saturated. }
\tablenotetext{b}{These eqwivalent widths were not listed in Y06}
\tablenotetext{c}{These lines are blended with lines of other species.}
\end{table}

\begin{table}
{\scriptsize
\caption{Relative \ebv $~$values for several sub-samples}
\bigskip
\begin{tabular}{|l|l|l|l|}
\tableline
\multicolumn{1}{|l|}{\bf Sample A}&\multicolumn{1}{|l|}{\bf Sample
B}&\multicolumn{1}{|l|}{\bf Description of comparison}&\multicolumn{1}{|l|}{\bf
$E(B-V)$\tablenotemark{a}}\\ 
\hline
250 RD, non-absorber QSOs&250 RUD, non-absorber QSOs& Non-absorber:
RD/RUD&0.036$\pm$0.0001\\
\#1&SY06&Associated/Intervening& 0.027$\pm$0.0001\\
\#10&\#11&Associated: RD/RUD&0.074$\pm$0.0002\\
SY06RD&SY06RUD&Intervening: RD/RUD&0.033$\pm$0.0001\\
\#10&SY06RD&RD: Associated/Intervening&0.062$\pm$0.0001\\
\#11&SY06RUD&RUD: Associated/Intervening&0.018$\pm$0.001\\
\#10 with \wmg $\ge$1.58&\#10 with \wmg $<$1.58&Associated, RD: strong Mg
II/weak Mg II&0.092$\pm$0.003\\
\#10 with 1.22$\le$\wmg $\le$2.1&\#10 with \wmg$<$1.22&Associated, RD:
intermediate Mg II/weak Mg II&0.074$\pm$0.001\\
\#10 with \wmg $\ge$2.1&\#10 with 1.22$\le$\wmg $<$2.1 &Associated, RD: strong
Mg II/intermediate Mg II&0.053$\pm$0.001\\
Color-selected QSOs &Color-selected QSOs &RD, color-selected:&\\
from sub-sample \#10& from SY06RD&Associated/Intervening&0.048$\pm$0.0001\\

\hline
\end{tabular}
}
\tablenotetext{a}{Relative $E(B-V)$ of sample A with respect to sample B,
obtained by fitting an SMC extinction curve to the ratio of composite spectra
for sub-sample A and B. To estimate the uncertainty of the $E(B-V)$ value, ten
simulated extinction curves were generated by adding noise to the original
extinction curve, consistent with the error array.  The values of $E(B-V)$ were
calculated for each of ten simulated extinction curves, and the rms dispersion
of the values is taken as the estimate of the $E(B-V)$ value uncertainty.}
\end{table}

\clearpage
\thispagestyle{empty}
\setlength{\voffset}{33mm}
\begin{table}
{\scriptsize
\caption{The full absorber and matching non-absorber samples}
\begin{tabular}{|l|l|l|l|l|l|l|l|l|l|l|l|l|l|}
\hline
\multicolumn{8}{|c|}{Absorber sample}&\multicolumn{6}{|c|}{Non-absorber sample}\\
\hline
Plate&Fiber&MJD&$z_{em}$&$\Delta(g-i)$&$z_{ab}$&$i$ mag&$\beta$&
Plate&Fiber&MJD&$z_{em}$&$\Delta(g-i)$&$i$ mag\\
\hline
0388& 289& 51793& 1.2940&  0.6225& 1.2925& 18.979 & 0.001 &   0760& 124& 52264&1.2920& -0.1023& 18.978 \\  
0651& 496& 52141& 1.3570&  0.0484& 1.3522& 18.467 & 0.002 &   1044& 455& 52468&1.3610&  0.2461& 18.472 \\ 
0390& 418& 51900& 1.0130&  0.7486& 1.0134& 18.843 & 0.000 &   1165& 131& 52703&1.0150&  1.0350& 18.844 \\ 
0753& 149& 52233& 1.7640&  0.2852& 1.7731& 17.885 &-0.003 &   0835& 271& 52326&1.7760&  0.0737& 17.801\\ 
0653& 558& 52145& 1.8570&  0.2630& 1.8430& 19.055 & 0.005 &   0583& 469& 52055&1.8510&  1.3632& 19.044 \\ 
0753& 034& 52233& 1.7080&  0.1664& 1.6870& 18.100 & 0.008 &   0302& 550& 51688&1.7240& -0.0054& 18.095 \\ 
0392& 528& 51793& 1.5400&  0.3629& 1.5498& 18.687 &-0.004 &   0763& 117& 52235&1.5370&  0.1036& 18.700\\ 
0392& 189& 51793& 1.0660&  0.0550& 1.0500& 18.741 & 0.008 &   0917& 393& 52400&1.0700&  0.0108& 18.751 \\ 
0418& 205& 51817& 1.2500&  0.1302& 1.2515& 18.595 &-0.001 &   0404& 127& 51812&1.2460& -0.0663& 18.600\\ 
0655& 389& 52162& 1.4720& -0.1273& 1.4553& 17.470 & 0.007 &   0448& 433& 51900&1.4750&  0.1756& 17.481 \\ 
0655& 239& 52162& 1.8810&  0.0453& 1.8566& 18.393 & 0.009 &   0619& 377& 52056&1.8690& -0.0409& 18.398 \\ 
0655& 536& 52162& 1.5250&  0.0190& 1.5230& 18.952 & 0.001 &   0812& 564& 52352&1.5220& -0.1874& 18.934 \\ 
0655& 177& 52162& 1.1630&  0.0058& 1.1425& 18.467 & 0.010 &   1163& 280& 52669&1.1630& -0.0206& 18.477 \\ 
0419& 056& 51879& 1.3840& -0.2027& 1.3699& 18.218 & 0.006 &   0634& 471& 52164&1.3780& -0.0937& 18.219 \\ 
0394& 242& 51913& 1.5810& -0.0437& 1.5559& 18.774 & 0.010 &   1287& 272& 52728&1.5810&  0.0727& 18.764 \\ 
0420& 155& 51871& 1.7300& -0.0071& 1.7027& 18.162 & 0.010 &   0924& 493& 52409&1.7200& -0.0002& 18.183 \\ 
0658& 527& 52146& 1.2720& -0.2076& 1.2754& 18.928 &-0.001 &   0751& 286& 52251&1.2720& -0.1203& 18.934\\ 
0397& 267& 51794& 1.3730&  0.1363& 1.3710& 17.248 & 0.001 &   1332& 519& 52781&1.3710& -0.0953& 17.200 \\ 
0660& 190& 52177& 1.6920& -0.0260& 1.7007& 17.914 &-0.003 &   1350& 297& 52786&1.6990& -0.1325& 17.850\\ 
0398& 110& 51789& 1.4970&  0.4441& 1.4864& 18.088 & 0.004 &   0520& 024& 52288&1.4990&  0.0729& 18.059 \\ 
0661& 362& 52163& 1.8360&  0.0305& 1.8078& 17.151 & 0.010 &   0326& 292& 52375&1.7700&  0.0479& 17.108 \\ 
0425& 447& 51898& 1.3040&  0.6804& 1.2921& 18.830 & 0.005 &   0749& 432& 52226&1.3090&  0.3725& 18.822 \\ 
0662& 565& 52147& 1.8390& -0.0588& 1.8181& 18.987 & 0.007 &   1171& 305& 52753&1.8400& -0.0192& 18.992 \\ 
0401& 255& 51788& 1.6590& -0.1123& 1.6746& 17.495 &-0.006 &   1209& 555& 52674&1.6640& -0.1576& 17.405\\ 
0429& 328& 51820& 1.7590& -0.0165& 1.7909& 18.186 &-0.011 &   0345& 128& 51690&1.7440&  0.0256& 18.193\\ 
0430& 362& 51877& 1.7380&  0.4237& 1.7385& 19.918 & 0.000 &   0453& 195& 51915&1.7350&  0.0215& 19.912 \\ 
0427& 402& 51900& 1.7000&  0.0348& 1.7085& 17.618 &-0.003 &   0448& 172& 51900&1.7070& -0.0999& 17.595\\ 
0404& 070& 51812& 1.5200&  0.0978& 1.5053& 19.077 & 0.006 &   0516& 108& 52017&1.5190& -0.1447& 19.070 \\ 
0667& 220& 52163& 1.7340&  0.4091& 1.7205& 18.902 & 0.005 &   1044& 230& 52468&1.7300& -0.1700& 18.929 \\ 
0405& 073& 51816& 1.5570& -0.1534& 1.5838& 18.089 &-0.010 &   0523& 062& 52026&1.5540&  0.0543& 18.047\\ 
0454& 064& 51908& 1.2540&  0.3680& 1.2565& 18.946 &-0.001 &   0761& 096& 52266&1.2520& -0.0956& 18.942\\ 
0455& 638& 51909& 1.2200& -0.1423& 1.2134& 19.833 & 0.003 &   0648& 238& 52559&1.2150&  0.7185& 19.830 \\ 
0410& 318& 51816& 1.4600& -0.0996& 1.4344& 17.738 & 0.010 &   0540& 568& 51996&1.4650&  0.1462& 17.772 \\ 
0457& 354& 51901& 1.4560& -0.0627& 1.4330& 18.178 & 0.009 &   0590& 282& 52057&1.4510&  0.3246& 18.178 \\ 
\hline
\end{tabular}
}
\end{table}
\clearpage
\setlength{\voffset}{0mm}

{\rotate{
\begin{table}
{\scriptsize
\caption{Equivalent widths of chosen lines\tablenotemark{a}}
\begin{tabular}{|l|l|l|l|l|l|l|l|l|l|l|l|}
\hline
Plate&Fiber&MJD&\multicolumn{9}{|c|}{Rest equivalent widths in {\AA}}\\
\cline{ 4 - 12 }
&&& Al II $\lambda1670$&C IV$\lambda1548$&C IV $\lambda1550$&Mg
I$\lambda2852$&Mg II $\lambda2796$&Mg II$\lambda2803$&Si IV $\lambda1393$&Si IV
$\lambda1403$&Fe II$\lambda2382$\\
\hline
0388& 289& 51793&-1.00,   0.34 & -1.00,  -1.00 & -1.00,  -1.00 &   0.68,   0.11 &  0.57,   0.09 & 0.56,   0.08 & -1.00,  -1.00 & -1.00,  -1.00 & -1.00,   0.17 \\  
0651& 496& 52141& 0.52,   0.12 & -1.00,  -1.00 & -1.00,  -1.00 &  -1.00,   0.20 &  0.98,   0.06 & 0.78,   0.05 & -1.00,  -1.00 & -1.00,  -1.00 &  0.62,   0.08 \\ 
0390& 418& 51900&-1.00,  -1.00 & -1.00,  -1.00 & -1.00,  -1.00 &  -1.00,   0.15 &  0.80,   0.09 & 0.70,   0.08 & -1.00,  -1.00 & -1.00,  -1.00 &  0.38,   0.12 \\ 
0753& 149& 52233& 1.23,   0.10 &  2.10,   0.10 &  1.27,   0.09 &   0.44,   0.12 &  2.60,   0.10 & 2.45,   0.08 &  1.00,   0.14 &  1.12,   0.15 &  1.44,   0.08\\ 
0653& 558& 52145& 0.98,   0.16 &  1.61,   0.11 &  1.02,   0.09 &  -1.00,   0.27 &  2.77,   0.20 & 1.51,   0.17 &  0.76,   0.16 & -1.00,   0.25 &  0.95,   0.12 \\ 
0753& 034& 52233&-1.00,   0.08 &  0.36,   0.06 &  0.19,   0.06 &  -1.00,   0.16 &  0.87,   0.09 & 0.54,   0.07 & -1.00,  -1.00 & -1.00,  -1.00 &  0.47,   0.06 \\ 
0392& 528& 51793& 0.93,   0.22 &  1.24,   0.20 &  1.11,   0.22 &   0.74,   0.11 &  1.96,   0.07 & 1.60,   0.07 & -1.00,  -1.00 & -1.00,  -1.00 &  1.23,   0.13\\ 
0392& 189& 51793&-1.00,  -1.00 & -1.00,  -1.00 & -1.00,  -1.00 &  -1.00,   0.23 &  2.32,   0.16 & 2.00,   0.14 & -1.00,  -1.00 & -1.00,  -1.00 &  1.20,   0.11 \\ 
0418& 205& 51817&-1.00,  -1.00 & -1.00,  -1.00 & -1.00,  -1.00 &   0.70,   0.10 &  1.58,   0.08 & 1.78,   0.09 & -1.00,  -1.00 & -1.00,  -1.00 &  1.27,   0.08\\ 
0655& 389& 52162& 0.46,   0.04 & -1.00,  -1.00 & -1.00,  -1.00 &   0.16,   0.03 &  1.05,   0.04 & 0.92,   0.03 & -1.00,  -1.00 & -1.00,  -1.00 &  0.42,   0.04 \\ 
0655& 239& 52162& 0.48,   0.08 &  0.36,   0.06 &  0.16,   0.04 &  -1.00,   0.18 &  1.23,   0.11 & 1.06,   0.10 & -1.00,   0.09 &  0.46,   0.10 &  0.74,   0.07 \\ 
0655& 536& 52162& 0.82,   0.20 &  1.36,   0.10 &  1.36,   0.11 &   0.41,   0.07 &  1.72,   0.07 & 1.25,   0.07 & -1.00,  -1.00 & -1.00,  -1.00 &  1.13,   0.13 \\ 
0655& 177& 52162&-1.00,  -1.00 & -1.00,  -1.00 & -1.00,  -1.00 &   0.19,   0.05 &  1.46,   0.08 & 1.06,   0.07 & -1.00,  -1.00 & -1.00,  -1.00 &  0.39,   0.06 \\ 
0419& 056& 51879& 0.42,   0.12 & -1.00,  -1.00 & -1.00,  -1.00 &   0.43,   0.10 &  1.60,   0.06 & 1.51,   0.06 & -1.00,  -1.00 & -1.00,  -1.00 &  0.90,   0.08 \\ 
0394& 242& 51913& 0.70,   0.12 & -1.00,   0.11 & -1.00,   0.11 &   0.42,   0.05 &  1.37,   0.08 & 1.10,   0.08 & -1.00,  -1.00 & -1.00,  -1.00 &  0.63,   0.07 \\ 
0420& 155& 51871& 0.18,   0.05 &  0.62,   0.05 &  0.50,   0.04 &   0.55,   0.13 &  0.53,   0.06 & 0.64,   0.09 & -1.00,  -1.00 & -1.00,  -1.00 & -1.00,   0.06 \\ 
0658& 527& 52146&-1.00,  -1.00 & -1.00,  -1.00 & -1.00,  -1.00 &   0.96,   0.18 &  2.25,   0.15 & 2.30,   0.11 & -1.00,  -1.00 & -1.00,  -1.00 &  1.84,   0.25\\ 
0397& 267& 51794& 0.42,   0.06 & -1.00,  -1.00 & -1.00,  -1.00 &   0.22,   0.03 &  0.55,   0.03 & 0.49,   0.03 & -1.00,  -1.00 & -1.00,  -1.00 &  0.42,   0.04 \\ 
0660& 190& 52177& 0.31,   0.07 &  1.92,   0.07 &  0.68,   0.05 &   0.40,   0.08 &  0.82,   0.06 & 0.81,   0.05 & -1.00,  -1.00 & -1.00,  -1.00 & -1.00,   0.06\\ 
0398& 110& 51789&-1.00,   0.10 &  1.37,   0.14 &  2.63,   0.44 &  -1.00,   0.14 &  0.44,   0.06 & 0.38,   0.06 & -1.00,  -1.00 & -1.00,  -1.00 & -1.00,   0.08 \\ 
0661& 362& 52163&-1.00,   0.04 &  0.61,   0.04 &  0.51,   0.03 &   0.26,   0.07 &  0.42,   0.07 & 0.36,   0.05 &  0.40,   0.08 &  0.23,   0.05 & -1.00,   0.04 \\ 
0425& 447& 51898& 0.69,   0.03 & -1.00,  -1.00 & -1.00,  -1.00 &   0.67,   0.09 &  1.73,   0.08 & 1.68,   0.07 & -1.00,  -1.00 & -1.00,  -1.00 &  0.94,   0.28 \\ 
0662& 565& 52147& 0.63,   0.14 & -1.00,   0.12 &  0.40,   0.10 &  -1.00,   0.33 &  1.64,   0.21 & 2.09,   0.20 &  0.60,   0.15 & -1.00,   0.21 &  1.30,   0.12 \\
\hline
\end{tabular}
}

\tablenotetext{a}{In column 4 to 12, the first number is the equivalent width,
while the second number is the 1 $\sigma$ error, in {\AA}, in the absorber rest
frame.  -1.0 for the equivalent width indicates that the line was not detected,
while -1.0 for the error means that the line was not covered by the SDSS
spectrum.}
\end{table}
}}
\clearpage
\begin{figure}[h]
\epsscale{0.5}
\plotone{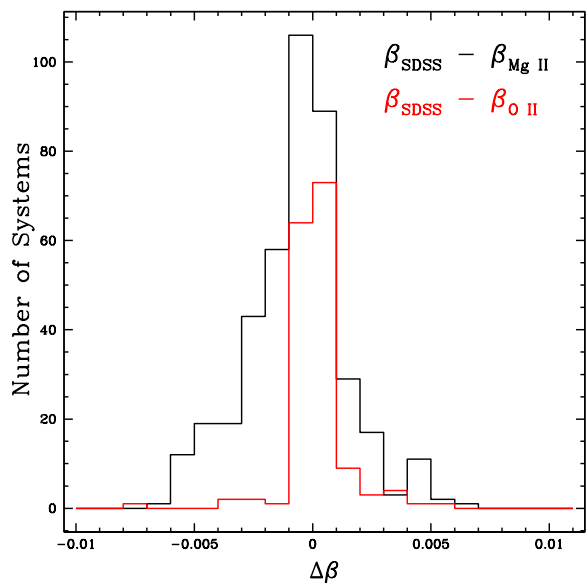}
\caption{The difference between the $\beta$ values derived for the Mg II
associated absorption systems using SDSS redshifts (\bet$_{\rm SDSS}$)and (red)
the single line [O II] (\bet$_{\rm O\;II}$) or (black) the single line Mg II
redshifts (\bet$_{\rm Mg\;II}$). [O II] is available for 162 QSOs, Mg II is
available for 415 QSOs. The single line [O II] and Mg II emission line
redshifts are derived from the SDSS pipeline. The [O II] line is regarded as
giving the best systemic redshift when it is available. The $\beta$ [O II]
values agree very well with the derived SDSS values, which are used in this paper.}
\end{figure}
\begin{figure}
\plottwo{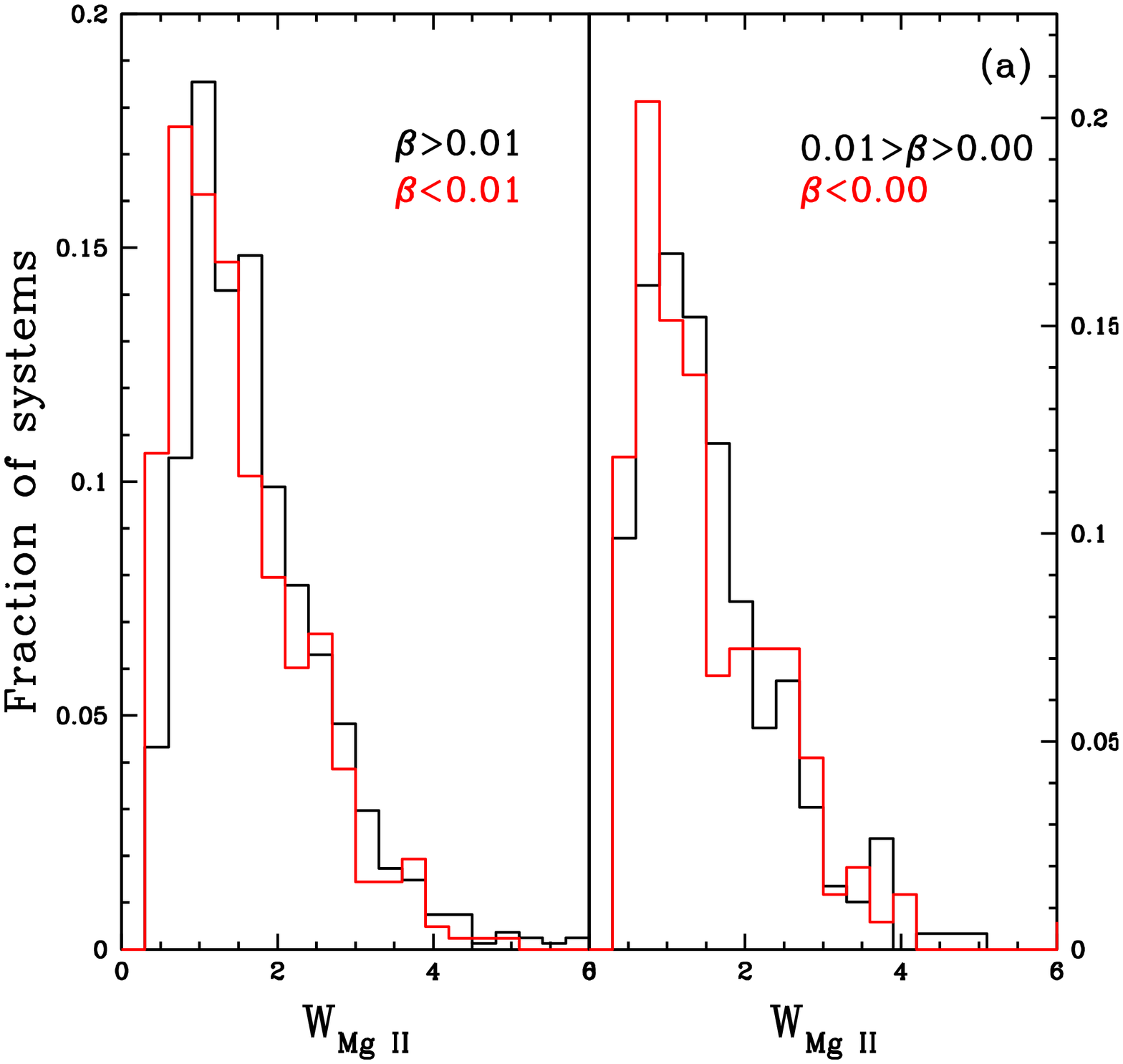}{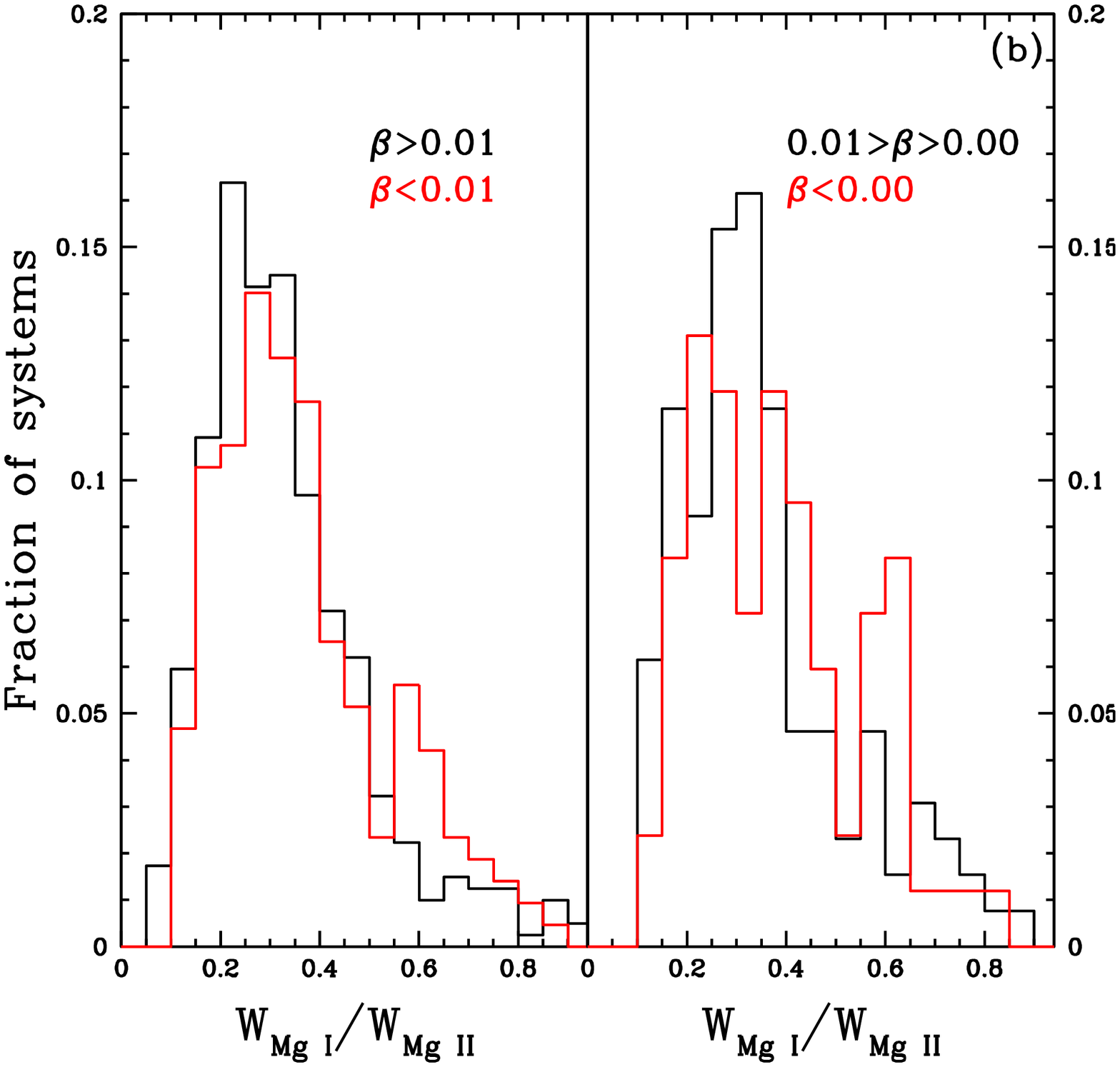}
\plottwo{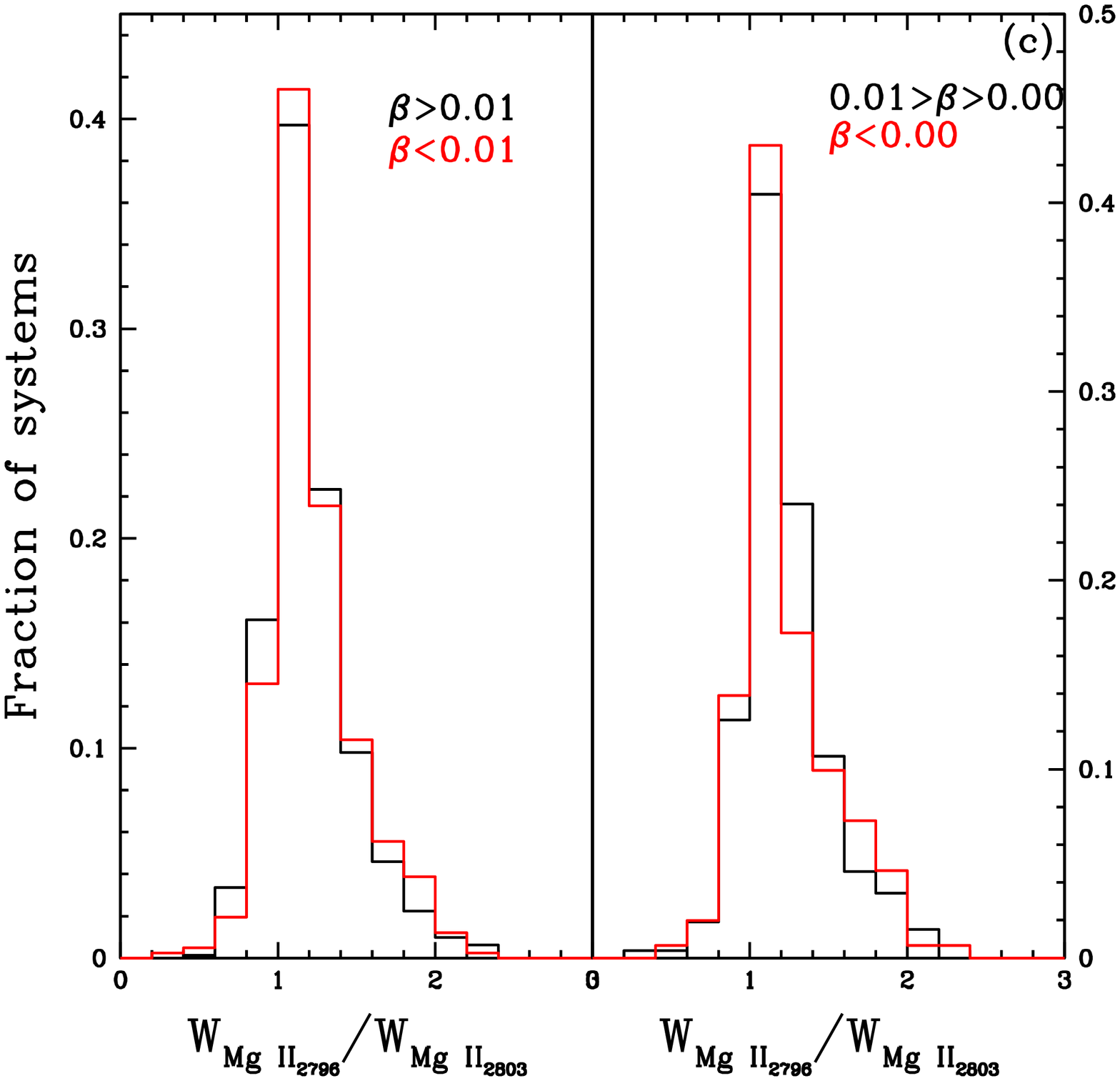}{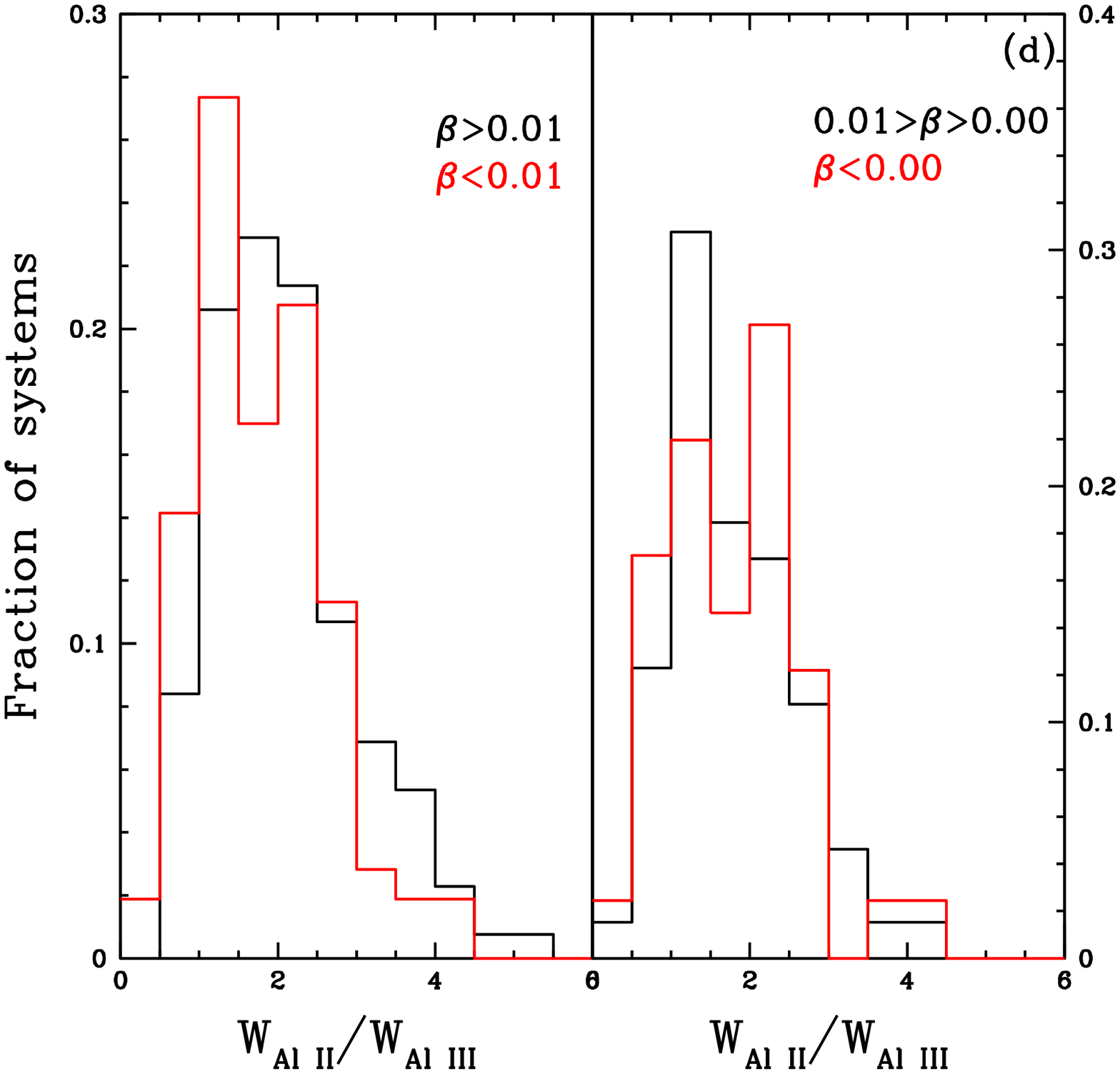}
\plottwo{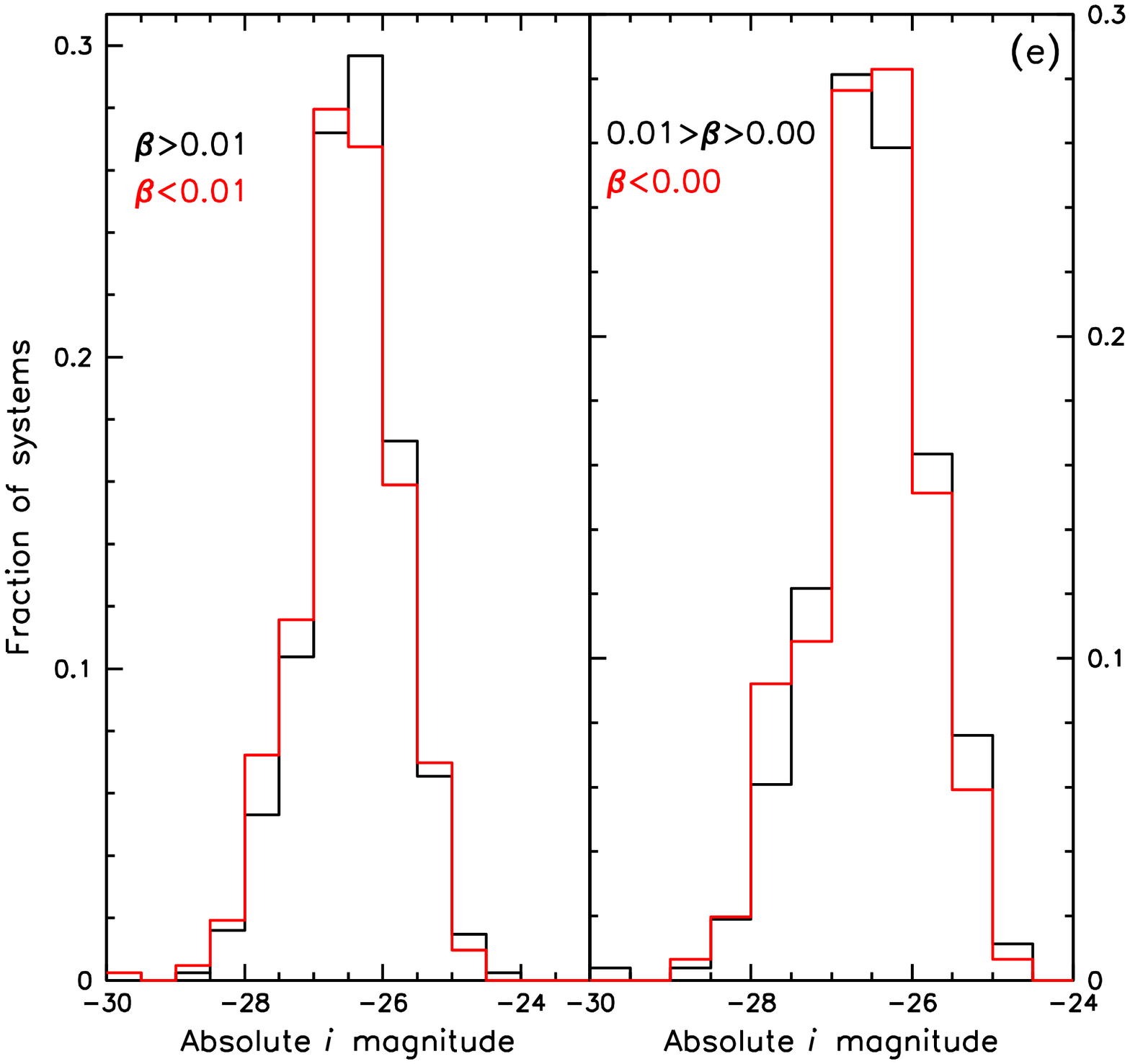}{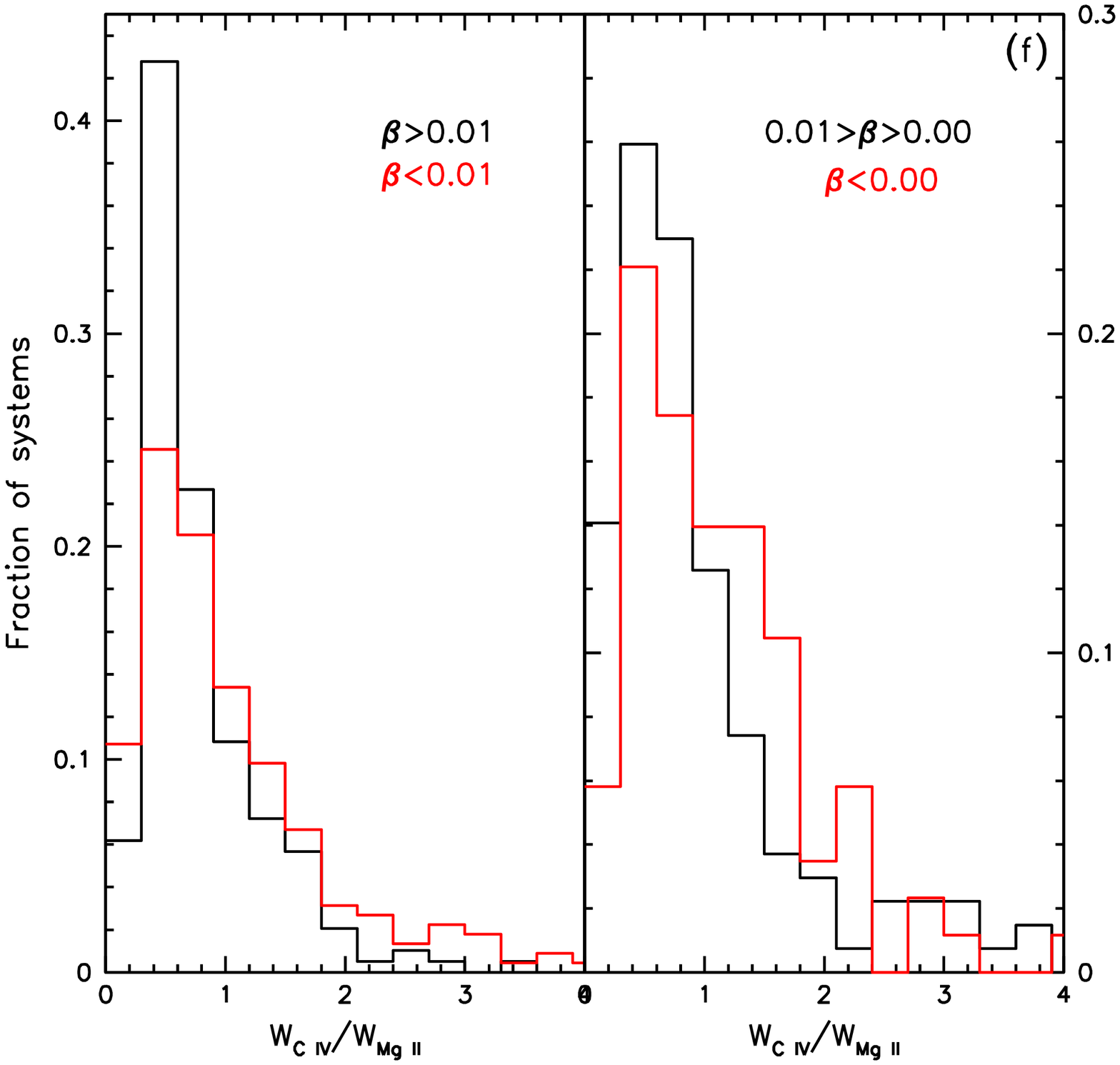}
\caption{Histograms comparing properties of associated (sample 1, this paper,
$\beta<$ 0.01, red) and intervening (SY06, $\beta>$ 0.01, black) systems appear
on the left side of five sub-panels which show (a) \wmg; (b) \wmgr; (c) the
{Mg~II} doublet ratio; (d) \walr; and (e) absolute $i$ magnitude. Panel (f)
shows the ratio W$_{\rm C\;IV}$/W$_{\rm Mg\;II}$, taken as an indicator of
ionization.  The left side of that panel compares the sample of associated
systems (sub-sample \#12, \bet$<$0.01, red) with the intervening systems from
Y06 (sub-sample SY06CIV, \bet$>$0.01, black), that is the sub-samples with
complete coverage of C IV doublets. The right side of panels (a) to (e) show
the positive (black) and negative (red) \bet $~$associated systems from sample
\#1 (sub-sample \#s 6 and 7) of this paper. The right side of the panel (f)
shows positive (sub-sample \#13, black) and negative (sub-sample \#14, red)
\bet $~$sub-samples for the {Mg~II} systems from the associated systems in
sub-sample \#12 of this paper (complete in C IV).}
\end{figure}
\begin{figure}
\epsscale{1.0}
\plotone{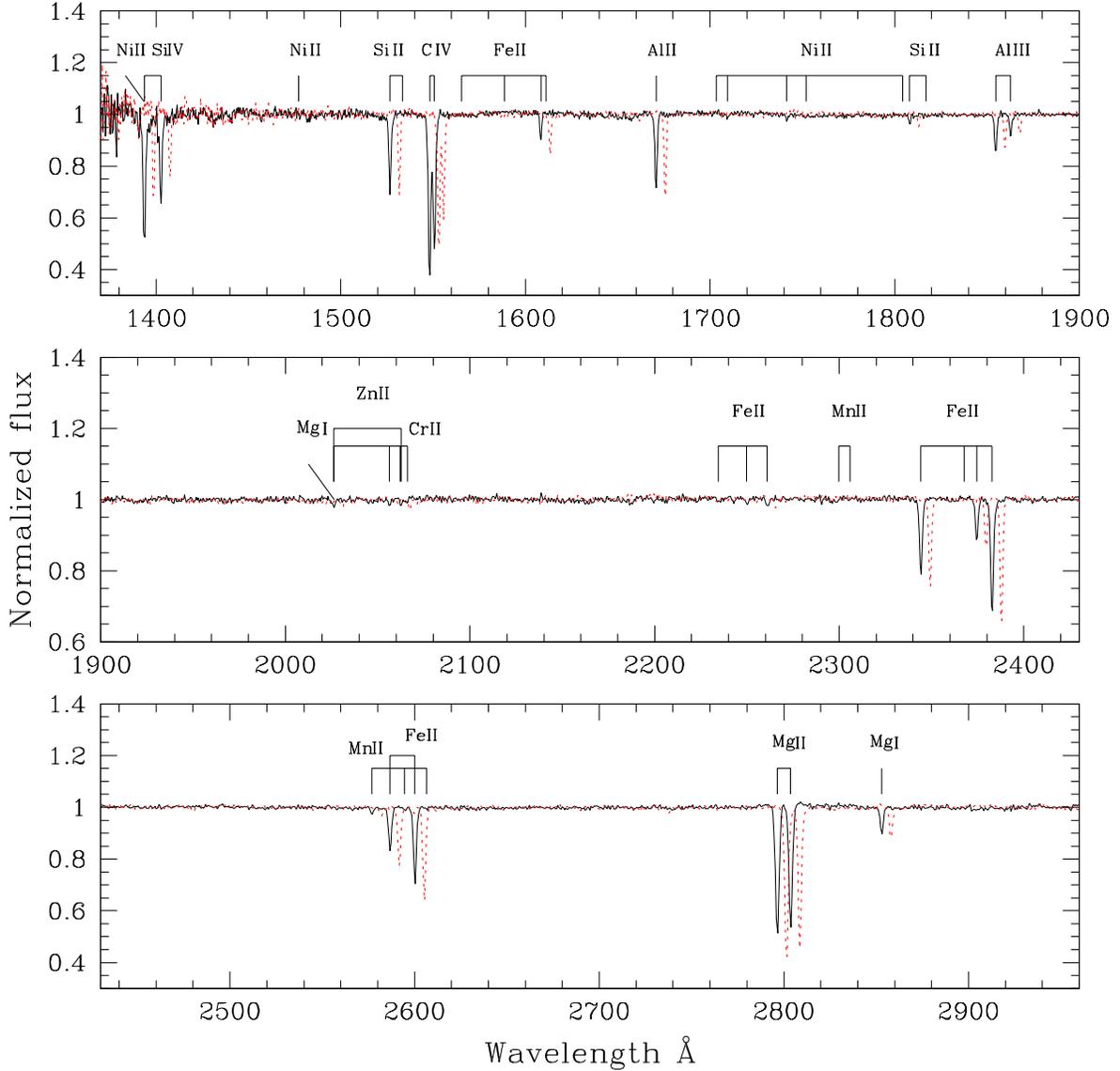}
\caption{The arithmetic mean composite spectrum of the systems in the
sub-sample (\#12) of associated systems  having \zab$>1.4675$ (so that the SDSS
spectrum includes the C IV doublet) in the absorber rest frame (black, solid
line).  Individual spectra have been normalized using the first 30 eigenspectra
of the principle components of the QSO spectra and weighted by inverse
variance. All spectra contribute to the region between 1540 and 3150 {\AA}
which, therefore, has high signal-to-noise.  The similar spectrum for
intervening absorbers (sub-sample SY06CIV, red, dotted line) culled from SY06
is shown for comparison. This spectrum is shifted by 5 {\AA} to the right for
ease in visual comparison. Please note the different y-axis scale in the middle
panel. The wavelength scale is uniform, so lines of a given velocity width
appear broader at longer wavelengths.}
\end{figure}
\begin{figure}
\plotone{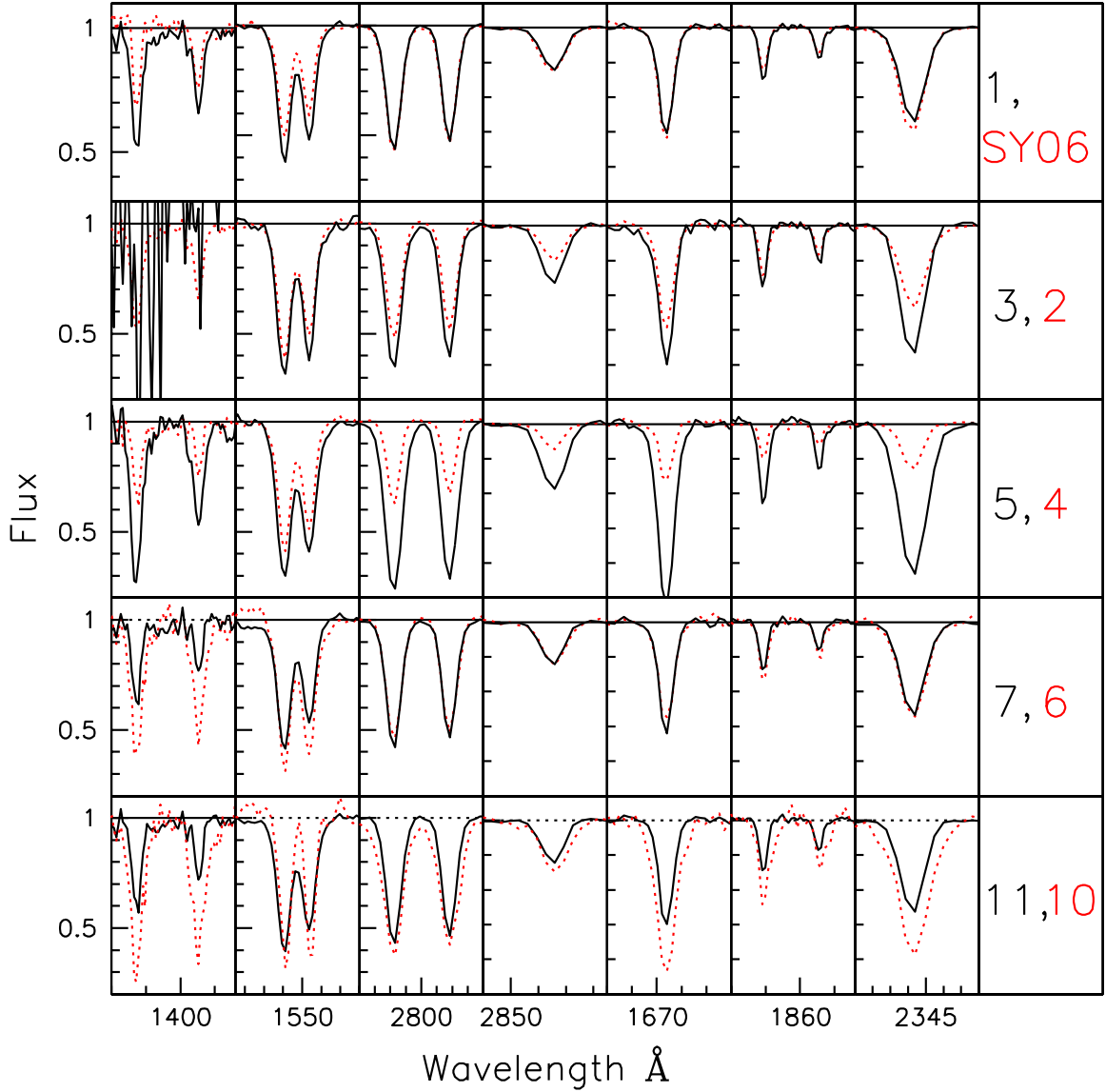}
\caption{Comparison of profiles of absorption lines Si IV $\lambda\lambda
1393,1402$, C IV $\lambda\lambda 1548,1550$, {Mg~II} $\lambda\lambda
2796,2803$, Mg I $\lambda$2852, Al II $\lambda$1670, Al III $\lambda\lambda$
1854,1864 and Fe II $\lambda$ 2344 in the arithmetic mean composite spectra of
sub-samples.  Sample numbers are given in the rightmost panels. Selection
criterion of the samples (as given in Table 1) are briefly described below for
ease in comprehension. The pairings are, with  the samples plotted and labeled
in red given first, SY06, \#1: full  samples of intervening and associated
absorbers; \#2, 3: \imag $<$ and $\ge$ 18.74; \#s 4, 5: \wmg $<$ and $\ge$ 1.35
{\AA}; \#s 6, 7: \bet $<$ and $\ge$ 0; \#s 10, 11: Radio-detected and
undetected QSOs.}
\end{figure}
\begin{figure}
\plotone{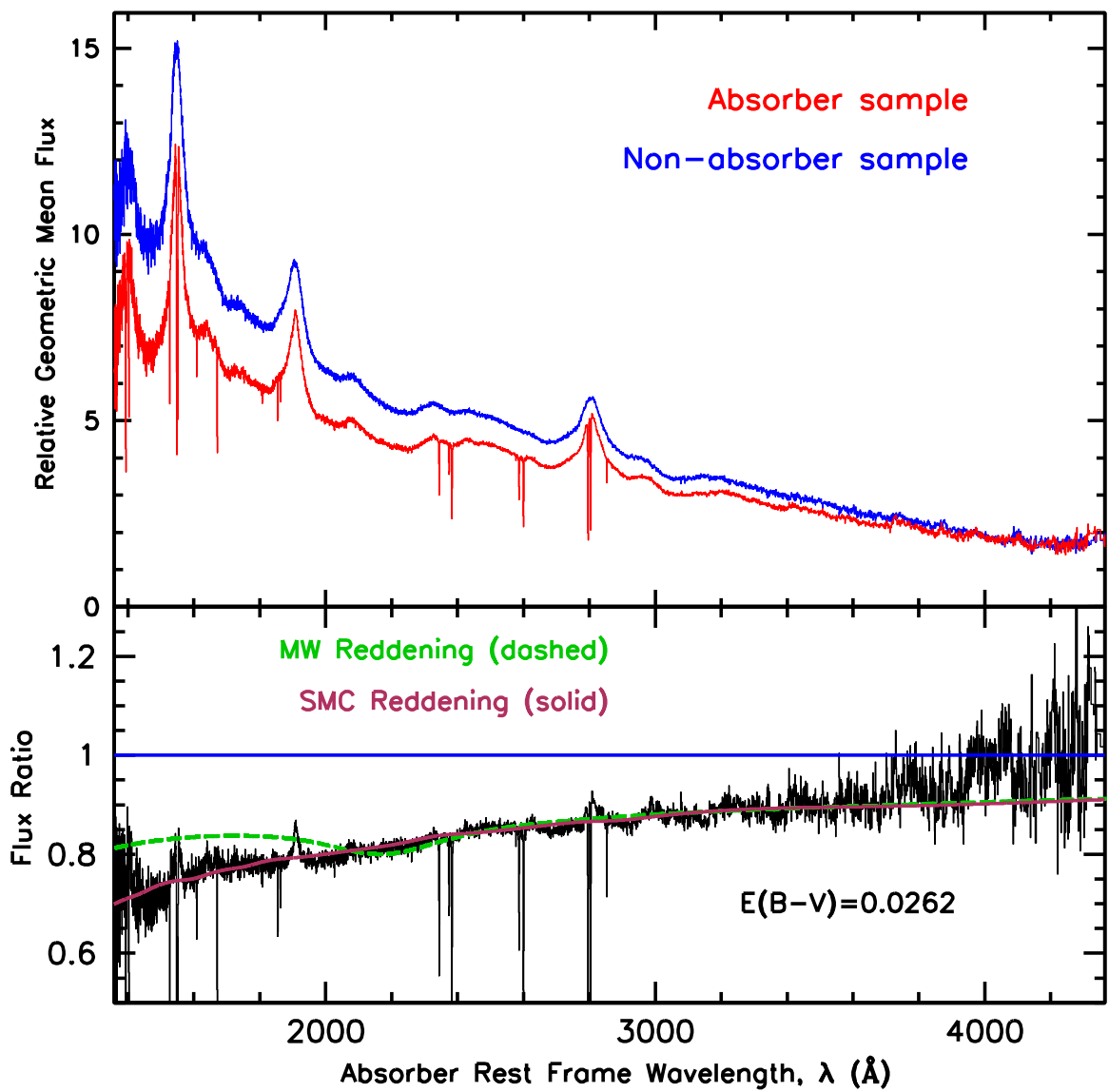}
\caption{The top panel shows the geometric mean composite spectrum of the QSOs
in sample \#1, in the absorber rest-frames, in red, and that of the
corresponding non-absorbers, in blue. Note that emission lines do appear in the
absorber rest-frame composite spectrum due to the small difference in emission
and absorption redshifts of the QSOs in the sample of associated systems. The
absorption lines in the red spectra are obvious. See text for a discussion of
the normalization of the plots. The bottom panel shows the ratio of the two
spectra along with the best fit SMC extinction curve in red.  The derived
absorber rest frame $E(B-V)$ value is given. The MW extinction curve for the
derived SMC $E(B-V)$ value is shown in green.}
\end{figure}
\begin{figure}[h]
\plottwo{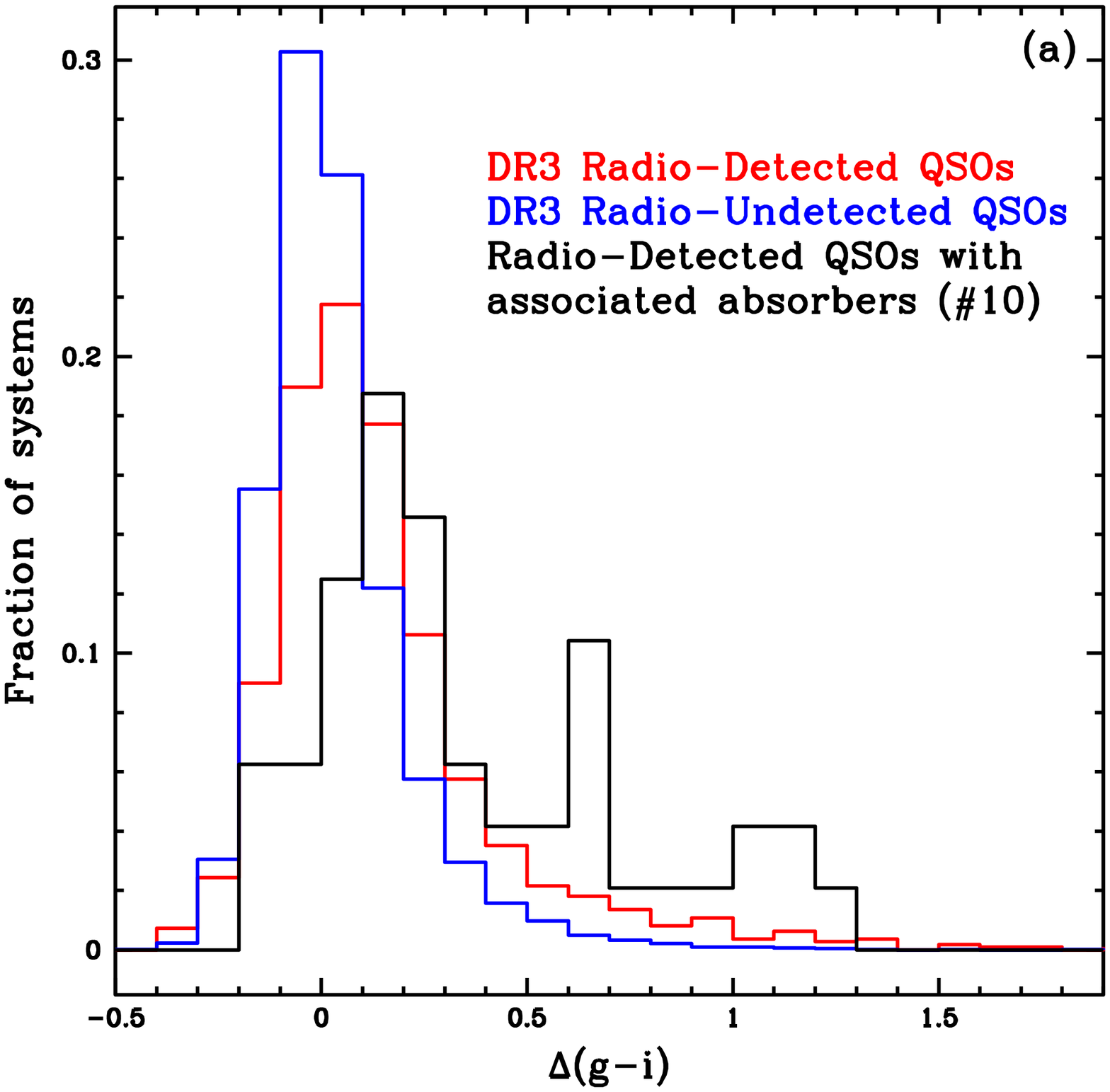}{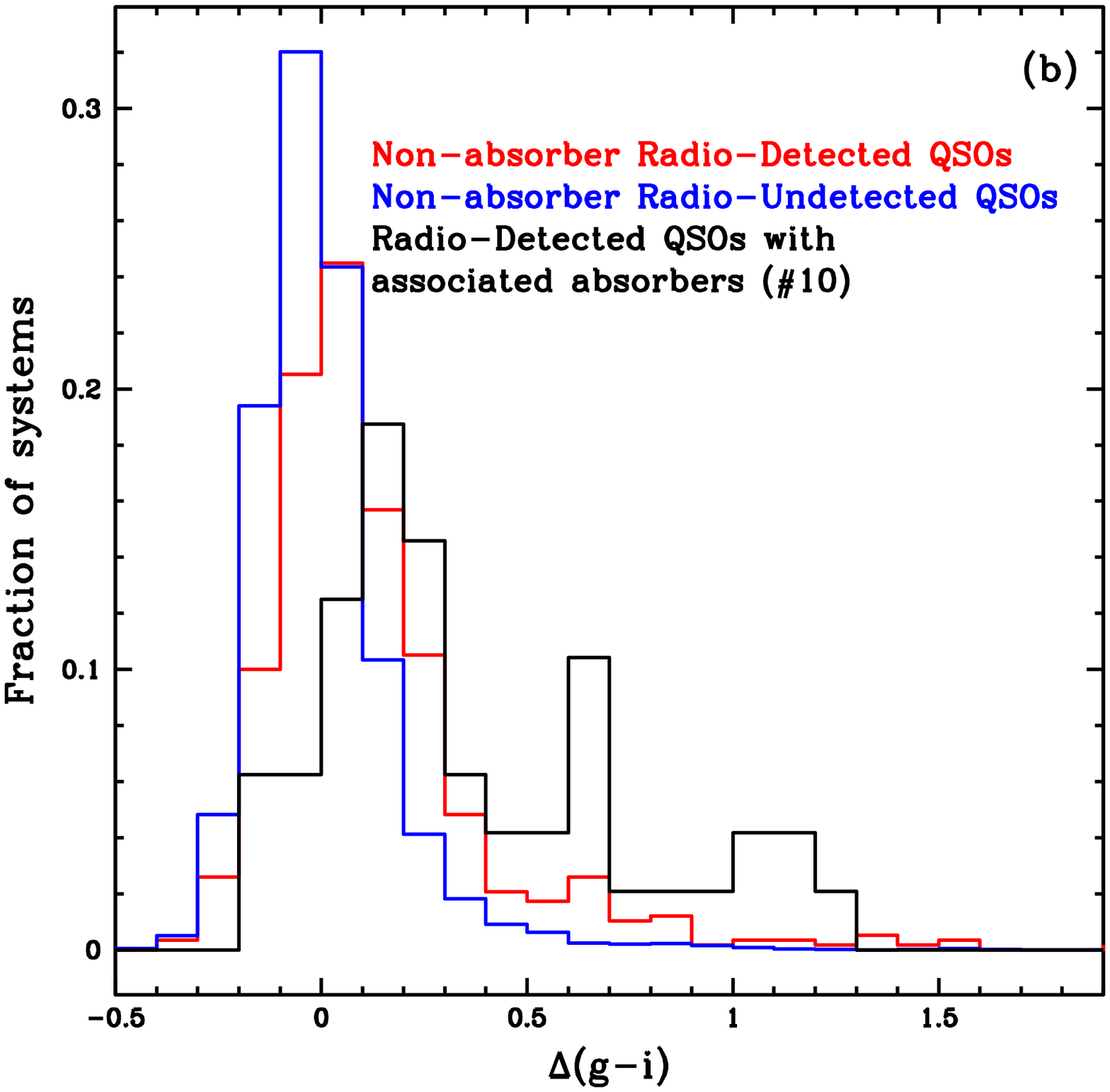}
\caption{The left panel (a) shows the distribution of \dgi $~$in SDSS DR3 QSOs
with $1.0<z_{em}<1.96$, having (i) non-zero FIRST flux (red) and (ii)
non-detection in the FIRST survey (blue).  Also shown is the distribution for
FIRST radio-detected QSOs in our sub-sample (\#10) of associated systems
(black). The right panel (b) shows the distribution of \dgi in SDSS DR3 QSOs
with $1.0<z_{em}<1.96$, confirmed by the present study as having no absorption
systems in their spectra and having (i) non-zero FIRST flux (red) and (ii)
non-detection in the FIRST survey (blue). Also shown is the distribution for
FIRST radio-detected QSOs in our sub-sample (\#10) of associated systems.}
\end{figure}

\end{document}